\documentclass[Afour,sagev,times,doublespace]{sagej}

\usepackage{color} 

\usepackage{subfig}

\usepackage[colorlinks,bookmarksopen,bookmarksnumbered,citecolor=red,urlcolor=red]{hyperref}

\newcommand\BibTeX{{\rmfamily B\kern-.05em \textsc{i\kern-.025em b}\kern-.08em
T\kern-.1667em\lower.7ex\hbox{E}\kern-.125emX}}

\def\volumeyear{2020}

\newcommand{\prova}{\par\noindent\textbf{Proof.} }
\newcommand{\fineprova}{\phantom{A}\hfill{$\blacksquare$}\goodbreak\medskip}

\newcommand{\loc}{\EVE}

\newcommand{\LLc}{\l_c}

\newcommand{\LLL}{\textrm{L}}

\newcommand{\kernel}{\phi}

\newcommand{\TANG}{T}

\newcommand{\conf}{{\BOmega}}

\newcommand{\di}[1]{(#1)}

\newcommand{\BOmega}{\boldsymbol{\Omega}}

\newcommand{\Bpi}{\boldsymbol{\pi}}

\newcommand{\Bss}{\boldsymbol{\sigma}}

\newcommand{\Bgamma}{\boldsymbol{\gamma}}

\newcommand{\Bn}{\mathbf{n}}
\newcommand{\Bo}{\mathbf{o}}
\newcommand{\Bp}{\mathbf{p}}

\newcommand{\Bu}{\mathbf{u}}
\newcommand{\Bv}{\mathbf{v}}
\newcommand{\Bw}{\mathbf{w}}

\newcommand{\BD}{\mathbf{D}}

\newcommand{\cM}{\mathcal{M}}

\newcommand{\VV}{V}

\newcommand{\LL}{L}

\newcommand{\TT}{T}

\newcommand{\MM}{M}
\newcommand{\NN}{N}

\newcommand{\scalar}[2]{{\langle}\kern.1em#1,#2\kern.1em{\rangle}}
\newcommand{\scalarC}[2]{{\langle}\kern.1em#1,#2\kern.1em{\rangle}_\conf}
\newcommand{\scalarT}[2]{{\langle}\kern.1em#1,#2\kern.1em{\rangle}_\TT}
\newcommand{\scalarTC}[2]{{\langle}\kern.1em#1,#2\kern.1em{\rangle}_{\TANG\conf}}

\newcommand{\integrale}[2]{\int_{#1}^{#2}}

\newcommand{\ee}{\varepsilon}

\newcommand{\sss}{s}
\newcommand{\mathscr}{\mathcal}

\newcommand{\sub}[1]{{}_{\lower2pt\hbox{$\scriptstyle#1$}}}

\newcommand{\punto}{\cdot}

\newcommand{\equaldef}{:=}

\newcommand{\der}{d}

\newcommand{\norma}[1]{\| #1 \|}

\numberwithin{theorem}{section}
\numberwithin{proposition}{section}
\numberwithin{lemma}{section}
\numberwithin{remark}{section}
\numberwithin{definition}{section}

\def\c{\textit{c}}
\def\s{\textit{s}}
\def\t{\textbf{t}}
\def\tort{\t_{\bot}}
\def\k{\textbf{k}}
\def\n{\textbf{n}}
\def\R{\textbf{R}}

\def\cinematica{\Bw}
\usepackage{graphicx}
\usepackage[demo]{adjustbox}
\usepackage{xcolor}
\usepackage{atbegshi}

\begin{document}
	\pagestyle{empty} 
	\begin{titlepage}
		\color[rgb]{.4,.4,1}
		\hspace{5mm}
		\hspace{15mm}
		\begin{minipage}{10mm}
			\color[rgb]{.7,.7,1}
			\rule{1pt}{226mm}
		\end{minipage}
		\begin{minipage}{133mm}
			\vspace{10mm}        
			\color{black}
			\sffamily
	\LARGE\bfseries Stress-driven two-phase integral elasticity for Timoshenko curved beams  \\[-0.3\baselineskip]   \\[-0.3\baselineskip] 
			
			\vspace{5mm}
			{\large {Preprint of the article published in \\[-0.4\baselineskip] Proceedings of the Institution of Mechanical Engineers, Part N: Journal of Nanomaterials, Nanoengineering and Nanosystems \\[-0.1\baselineskip] 235(1-2), February 2021, 52-63}}  
			\vspace{10mm}        
			{\large \\[-0.4\baselineskip] Marzia Sara Vaccaro,\\[-0.4\baselineskip] Francesco Paolo Pinnola, \\[-0.4\baselineskip]  Francesco Marotti de Sciarra,\\[-0.4\baselineskip] Marko Canadija, \\[-0.4\baselineskip] Raffaele Barretta} 
			
			\large
			
			\vspace{40mm}
			\vspace{5mm}
			
			\small
			\url{https://doi.org/10.1177/2397791421990514}
			
			\textcircled{c} 2021. This manuscript version is made available under the CC-BY-NC-ND 4.0 license \url{http://creativecommons.org/licenses/by-nc-nd/4.0/}
			\hspace{30mm} 
			\color[rgb]{.4,.4,1} 
		\end{minipage}
	\end{titlepage}


\runninghead{Vaccaro, Pinnola, Marotti de Sciarra, Canadija and Barretta}

\title{Stress-driven two-phase integral elasticity for Timoshenko curved beams}

\author{Marzia S Vaccaro\affilnum{1}, Francesco P Pinnola\affilnum{1},
Francesco Marotti de Sciarra\affilnum{1}, Marko Canadija\affilnum{2}, Raffaele Barretta\affilnum{1}
}

\affiliation{\affilnum{1}Department of Structures for Engineering and Architecture, 
		University of Naples Federico II, Naples, Italy\\
\affilnum{2}Department of Engineering Mechanics,
	   Faculty of Engineering, University of Rijeka, Rijeka, Croatia}

\corrauth{Raffaele Barretta\\
Department of Structures for Engineering and Architecture\\
		University of Naples Federico II,
		via Claudio 21, 80125 - Naples, Italy.}

\email{rabarret@unina.it}

\begin{abstract}
In this research, the size-dependent static behaviour of elastic curved stubby beams is investigated by Timoshenko kinematics.
Stress-driven two-phase integral elasticity is adopted to model size effects which soften or stiffen classical local responses.
The corresponding governing equations of nonlocal elasticity are established and discussed, non-classical boundary conditions are detected and an effective coordinate-free solution procedure is proposed.
The presented mixture approach is elucidated by solving simple curved small-scale beams of current interest in Nanotechnology.
The contributed results could be useful for design and optimization of modern sensors and actuators.
\end{abstract}


\maketitle

\section{Introduction}
\label{sec: intro}
Analysis and modelling of scale phenomena in micro- and nano-structures is a subject of current interest in Engineering Science \citep{Farajpour2018}.
Development of simple and computationally convenient methodologies for design and optimization of modern devices 
\citep{Tran2018,Basutkar2019a,Caplins2019,Ogi2019,Natsuki2019,Basutkar2019b,Ghayesh103202}
and nanocomposites 
\citep{Pourasghar2019,Eyvazian2020,Omari2020}
has been the main motivation of numerous investigations.
Crucial point is to take in due account small-scale effects which are technically significant and cannot be overlooked 
\citep{RafiiTabar2016}.
Assessment of size effects can be advantageously performed by making recourse to tools and techniques of nonlocal continuum mechanics rather than time consuming atomistic approaches
\citep{GhavanlooRF2019}.
Seminal treatments on nonlocal theory of elasticity, based on integro-differential formulations, were mainly conceived to be applied to engineering problems involving dislocations and waves \citep{Rogula1965,Rogula1976,Rogula1982,Eringen1972}.
Basic concepts of nonlocal mechanics can be consulted in the paper by \citet{Bazant2002}.
Strain-driven integral formulations were reverted by \citet{Eringen1983} to more convenient sets of differential equations due to tacit and rapid vanishing of nonlocal fields governing relevant convolutions defined in unbounded domains.
Such a mathematical scenario has been recently proven to be not permissible if strain-driven integral equations of pure elasticity are exploited in order to describe size effects in small-scale structures of technical interest. 
This issue has been comprehensively discussed by \citet{Romano2017} and recently acknowledged by the scientific community, see e.g. \citep{Sidhardh2020,ZhangCS2020,Dilena2019,Shahsavari2018,Vila2017}.
Several proposals have been contributed in literature to bypass the apparent conflict between equilibrium and non-classical constitutive boundary conditions associated with the strain-driven integral convolution.
An example is the integral approach illustrated and applied  to nanorods by \citet{Maneshi2020}, framed in the research field of nonlocal models providing compensation of boundary effects \citep{Pijaudier1987,Borino2003,Polizzotto2004,Koutsoumaris2017,Fuschi2019}. 
As applied by \citet{Polizzotto2001}, Eringen's integral approach can be amended by considering a strain-driven two-phase (local/nonlocal) mixture \citep{Eringen1987} which leads to well-posed elastostatic problems, provided that the local contribution is not vanishing
\citep{RomanoIntModels2017}. 
The strain-driven mixture has been exploited by various authors \citep{Khodabakhshi2015,Wang2016,Fern2017} to model straight structures in several papers and recently by \citet{ZhangZAMM2019} to study circular curved beams. 
Total remedy to difficulties and singularity of Eringen's formulations in structural mechanics are overcome if the stress-driven integral methodology \citep{RB2017} is adopted for modelling size effects.
The well-posed approach has been recently extended by \citet{BarrettaPhysE2018} to capture both softening and stiffening elastic responses characterizing small structural scales by proposing for straight beams a stress-driven two-phase (local/nonlocal) mixture which is non-singular for any local fraction.

Motivation of the present paper is to generalize the aforementioned stress-driven mixture formulation to stubby curved beams which are basic structural components of modern micro- and nano-systems.
The proposed methodology is valid for arbitrary geometry of beam axis and provides the following special treatments involving:
\begin{enumerate}
\item
stress-driven integral slender (Bernoulli-Euler) beams \citep{BarrettaCurved2019};
\item
stress-driven integral stubby (Timoshenko) beams with uniform geometric curvature of the structural axis \citep{ZhangCS2020}.
\end{enumerate}
While the strategies contributed by \citet{BarrettaCurved2019,ZhangCS2020} are able to capture hardening structural behaviours for increasing nonlocal parameter, the approach developed in the present paper is conveniently able to model also softening elastic responses.
Accordingly, the proposed model can be applied to a wider class of applicative problems in nanomechanics. 
The plan is the following.
Timoshenko kinematic assumptions and ensuing equilibrium equations of curved beams are provided in the next section 
using a coordinate-free variational approach.
Classical local constitutive equations of elasticity for stubby beams are recalled and extended in the section named:
"Stress-driven mixture model" 
to tackle two-phase local/nonlocal materials.
The associated elastostatic problem is formulated and tackled by a simple solution procedure illustrated in the section entitled:
"Elastic equilibrium of Timoshenko curved nanobeams".
Selected case-studies of nanotechnological interest are then investigated and discussed.
Closing remarks are outlined in the last section.

\section{Kinematics and equilibrium of Timoshenko curved beams}\label{BEcin}

Let us consider a Timoshenko curved beam of length $L$ and denote by $\Bgamma$ a regular curve of the plane $\,\Bpi\,$ describing the beam axis and parameterized by the curvilinear abscissa $s \in [0, L]$.

\begin{figure}[h]
\centering	
\includegraphics[width=0.5\textwidth]{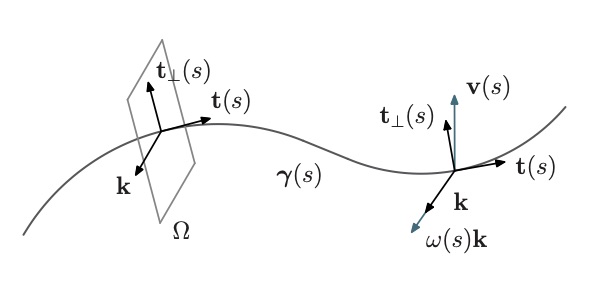}
\caption{Sketch of a Timoshenko beam with local triad and kinematic parameters.}
\label{ske}
\end{figure}

%


At each point of the axis is associated a cross-section modeled by a two-dimensional domain $\Omega$. 
Denoting by $\,(\bullet)'\,$ the derivative with respect to $s$, the tangent unit vectors field is defined as
\begin{equation}
\t \equaldef \Bgamma' 
\label{t}
\end{equation}

By applying the orthogonal linear transformation $\,\R\,$, which performs the rotation
by $\pi/2$ counterclockwise in the plane $\,\Bpi\,$, the transversal unit vectors field is obtained as

\begin{equation}
\tort\equaldef\R\t
\end{equation}

Then, we introduce the uniform unit vectors field $\,\k\equaldef\t\times\tort\,$.

Since the curve $\Bgamma$ is assumed to be regular, the vector $\t'$ is defined at each abscissa $s$. 
Thus, introducing the scalar geometric curvature of the beam axis $\,c\equaldef\norma{\t'}\,$, 
the normal unit vector is defined by $\,\Bn\equaldef\t'/c\,$. Vectors $\,\tort\,$ and $\,\Bn\,$ are related as \citep{RomTomoUno}  
$\,\tort=(\tort\punto\Bn)\,\Bn\,$.

The kinematic hypothesis of the Timoshenko beam model is that cross-sections are hinged to the beam axis;
thus, denoting by $\,\VV\,$ the linear space of translations, the velocity $\,\Bv:[0,\LLL]\mapsto\VV\,$ of the beam axis and the angular velocity of cross sections $\,\omega:[0,\LLL]\mapsto\Re\,$ are independent parameters describing the beam kinematics.

The tangent deformation field kinematically compatible with $\cinematica = \{\Bv,\:\omega\}$
is composed of axial strain, shear strain and flexural curvature scalar fields, i.e. $\{\ee_{\cinematica}, \gamma_{\cinematica}, \chi_{\cinematica}\} :[0,L]\mapsto\Re$.
\begin{equation}
\BD_{\cinematica}
=
\begin{vmatrix}
\ee_{\cinematica} \\
\gamma_{\cinematica}  \\
 \chi _{\cinematica}
\end{vmatrix}
=
\begin{vmatrix}
\Bv'\punto\t \\
(\Bv'\punto\tort) - \omega  \\
\omega'
\end{vmatrix}
\label{def}
\end{equation}



By duality with the tangent deformation field $\BD_{\cinematica}$, the stress $\,\Bss\,$ in a Timoshenko beam
is composed of axial force, shear force and bending moment scalar fields, that is $\{N, T, M\}:[0,L]\mapsto\Re\,$.

\begin{equation}
\Bss=
\begin{vmatrix}
N \\
T \\
M
\end{vmatrix}
\end{equation}

Equilibrium is expressed by the variational condition that the external virtual power of the force system $\textbf{\textit{l}}$ is equal to
the internal virtual power of the stress $\,\Bss\,$, for any virtual translation velocity and 
angular velocity, i.e.~$\delta\cinematica =  \{\delta\Bv,\:\delta\omega\}$, fulfilling homogeneous kinematic boundary conditions

\begin{equation}
\setlength{\jot}{8pt}
\begin{aligned}
\scalar{\textbf{\textit{l}}}{\delta\cinematica}
=
\integrale{0}{L}
\Bss\di\sss
\punto
\BD_{\delta\cinematica}\di\sss\,\der\sss
\,
\label{fm:VWC}
\end{aligned}
\end{equation}

%

Integrating by parts Eq.~\eqref{fm:VWC} we get vector differential equation and boundary conditions ruling equilibrium
between stress $\,\Bss\,$ and external force system $\textbf{\textit{l}}$ which is composed of a distributed vector loading $\,\Bp:[0,L]\mapsto\VV\,$, distributed bending couples $\,m:[0,L]\mapsto\Re\,$, boundary concentrated forces $\,{\mathbf{F}}_0\in\VV\,$ and $\,{\mathbf{F}}_L\in\VV\,$ and boundary concentrated bending couples $\,\cM_0\in\Re\,$ and $\,\cM_L\in\Re\,$.

Then, projecting the vector differential equation of equilibrium along $\t$ and $\tort$ directions we get the following 
differential problem in $\,[0,\LLL]\,$
\begin{equation}
\setlength{\jot}{8pt}
\left\{
\begin{aligned}
&\NN'-\TT\,\tort\punto\t'
=-\Bp\punto\t\,
\\
&\TT'-\NN\,\t\punto\tort'
=-\Bp\punto\tort\,
\\
&\TT + M' = - m
\label{fm:DiffProbl}
\end{aligned}
\right.
\end{equation}
equipped with the boundary conditions
at $\,\sss=0\,$ and $\,\sss=\LL\,$
\begin{equation}
\setlength{\jot}{8pt}
\left\{
\begin{aligned}
-(\NN\t+\TT\tort)\di{0}\punto\delta\Bv\di{0}
&={\mathbf{F}}_0\di{0}\punto\delta\Bv\di{0}\,
\\
(\NN\t+\TT\tort)\di{L}\punto\delta\Bv\di{L}
&={\mathbf{F}}_0\di{L}\punto\delta\Bv\di{L}\,
\\
-\MM\di{0}\;\delta\omega\di{0}
&=\cM_0\;\delta\omega\di{0}\,
\\
\MM\di{L}\;\delta\omega\di{\LL}
&=\cM_L\;\delta\omega\di{\LL}\,
\label{fm:StatBC}
\end{aligned}
\right.
\end{equation}
From Eqs.~\eqref{fm:StatBC}, natural static boundary conditions follow from essential kinematic boundary conditions related to the assigned constraints.
When essential boundary conditions are not prescribed, i.e. virtual translation velocity $\,\delta\Bv\,$ and 
angular velocity $\,\delta\omega\,$ are arbitrary, the corresponding natural static conditions from Eqs.~\eqref{fm:StatBC}
are:
\begin{equation}
\setlength{\jot}{8pt}
\left\{
\begin{aligned}
-(\NN\t+\TT\tort)\di{0}
&={\mathbf{F}}_0\di{0} \,
\\
(\NN\t+\TT\tort)\di{L}
&={\mathbf{F}}_0\di{L} \,
\\
-\MM\di{0}
&=\cM_0 \,
\\
\MM\di{L}
&=\cM_L
\label{fm:Stat}
\end{aligned}
\right.
\end{equation}
Explicit prescriptions of essential and natural boundary conditions for usual constraints adopted in structural mechanics are provided in the section named ''Case-studies''.

\section{Stress-driven mixture model}
\label{sec:StressDri}

\def\Det{\Lambda}
\def\loc{l}
\def\J{J_c}

In this section we will first recall the classical theory of local elasticity for a Timoshenko curved beam. Let us denote by $\,E\,$ and $\,G\,$
 Euler-Young and shear moduli, respectively.
$\,A\,$ stands for cross-sectional area and
$\,\J\,$ is the moment of inertia along the bending axis $\eta$ which is
identified by the direction of the transversal unit vector $\,\tort\,$, that is:

\begin{equation}
\J
=
\integrale{\Omega}{}\eta^2\frac{1}{1-c\,\eta \:(\n \cdot \tort)}\,\der A\,
\label{fm:RedMomIner}
\end{equation}

The constitutive equations of local elasticity for curved beams are expressed 
following the treatments by \citet{Winkler1858,Baldacci}, generalized by the 
presence of $\n \cdot \tort$, that is

\begin{equation}
\setlength{\jot}{8pt}
\left\{
\begin{aligned}
\ee_{\loc}\di{s} &= \frac{1}{EA}\biggr[N + \frac{\c\, M}{\n \cdot \tort}\biggr]\di{s}   \,
\\
\chi_{\loc}\di{s} &= 
\frac{M}{E\J}(s) + (\n \cdot \tort) \frac{\c}{EA}\biggr[N + \frac{ \c\, M}{\n \cdot \tort}\biggr]\di{s}\,
\\
\gamma_{\loc}\di{s} &= \bigg{[}\frac{T}{GK_r}\bigg{]}\di{s}
\end{aligned}
\right.
\label{LocEq9}
\end{equation}
where the equilibrium differential equation $\TT + M' = - m$ has been considered in Eq.~\eqref{LocEq9}$_3$, assuming vanishing distributed bending couples.

$K_r$ in Eq.~\eqref{LocEq9}$_3$ is the shear stiffness for a curved beam defined as follows


\begin{equation}
K_r^{-1} = \integrale{\Omega}{} \frac{r^2}{{[r - \eta(\n \cdot \tort)]}^2 \: b^2} \bigg{[} \frac{S_c}{\J} - \frac{A^*}{(\n \cdot \tort)rA}\bigg{]}^2  \,\der A
\label{K_r}
\end{equation}

where $r = 1/c$ is the curvature radius and $S_c$ is the static moment of $A^*$, that is

\begin{equation}
S_c
\equaldef
\integrale{\Omega^*}{} \frac{\eta}{1-c\,\eta \:(\n \cdot \tort)}\,\der A^* 
\label{fm:mom_statico_r}
\end{equation}

\def\f{\textbf{f}}
\def\s{\textbf{i}}

For a Timoshenko curved nanobeam the nonlocal elastic deformation fields are expressed by the stress-driven mixture model 
consisting in a convex combination of the local response in Eq.~\eqref{LocEq9} and a nonlocal response obtained by a convolution between
the local field and a scalar averaging kernel $\,\kernel :\Re \mapsto [0;+\infty[$.

Let preliminarily inticate with $\s$ and $\f$ the vectors collecting source and output fields, i.e.: $\s = \{\ee_{\loc}, \chi_{\loc}, \gamma_{\loc}\};\: \f = \{\ee, \chi, \gamma\}$.
Then, by denoting with $0 \leq m \leq 1$ the mixture parameter and with $l_c$ a positive nonlocal length parameter, the stress-driven mixture model is expressed as follows
\begin{equation}
\f\di{s}= m\: \s\di{s} + (1 - m) \: \int_{0}^{L}{\kernel_{\LLc}(s, \xi) \: \s\di{\xi}\:d\xi}    
\label{StressConv12}
\end{equation}
which is a Fredholm integral equation of the second kind in the unknown source field \s.

The averaging kernel is assumed to be the special bi-exponential function adopted by \citet{Eringen1983}, that is

\begin{equation}
\kernel_{\LLc}\di{s} = \frac{1}{2\LLc} \exp(-\frac{|s|}{\LLc})\,
\label{kern}
\end{equation}

By extending to Timoshenko curved beams the \textit{Mixture equivalence} by \citet[][Lemma 2]{RomanoIntModels2017}, it can be proved that the integral convolutions in Eq.~\eqref{StressConv12} is equivalent to the following differential equation
\begin{equation}
\frac{\f\di{s}}{\LLc^2} -\,\partial_s^2\f\di{s}
=\frac{\s\di{s}}{\LLc^2} - m\:\partial_s^2\s\di{s}
\label{EqDiff}
\end{equation}
equipped with the constitutive boundary conditions

\begin{equation}
\setlength{\jot}{8pt}
\left\{
\begin{aligned}
&\partial_s \f\di{0}=\frac{1}{\LLc}\,\f\di{0} + m\: \bigg{(} \partial_s \s\di{0} - \frac{\s\di{0}}{\LLc}\bigg{)} \,
\\
&\partial_s \f\di{L}=-\frac{1}{\LLc}\,\f\di{L} + m\: \bigg{(} \partial_s \s\di{L} + \frac{\s\di{L}}{\LLc}\bigg{)}      
\end{aligned}
\right.
\label{BCS}
\end{equation}

The stress-driven mixture model in Eq.~\eqref{StressConv12} is purely nonlocal when $m = 0$ and purely local when $m = 1$.
Since the model is based on two parameters (i.e. mixture and nonlocal length parameters) it is able to provide softening or stiffening
responses as shown in the section "Case-studies" and therefore can 
effectively model a wide range of nano-engineering problems.

\section{Elastic equilibrium of Timoshenko curved nanobeams}
\label{sec:elasticp}

\def\Bu{\textbf{u}}
\def\Bw{\textbf{w}}

Let us consider the linearized, plane and curved Timoshenko beam model and denote by $\,\Bu:[0,\LLL]\mapsto\VV\,$ the displacement vector field of the structural axis
and by $\,\varphi:[0,\LLL]\mapsto\Re\,$ the scalar field of rotations of cross-sections.

From Eq.~\eqref{def}, the kinematically compatible deformation field $\,\BD_\Bw\,$ associated with $\Bw = \{\Bu, \varphi\}$
is composed of axial strain $\ee_{\Bw}:[0,L]\mapsto\Re$,  shear strain $\gamma_{\Bw}:[0,L]\mapsto\Re$ and flexural curvature $\chi_{\Bw}:[0,L]\mapsto\Re$ scalar fields.

\begin{equation}
\BD_{\cinematica}
=
\begin{vmatrix}
\ee_{\cinematica} \\
\gamma_{\cinematica}  \\
 \chi _{\cinematica}
\end{vmatrix}
=
\begin{vmatrix}
\Bu'\punto\t \\
(\Bu'\punto\tort) - \varphi  \\
\varphi'
\end{vmatrix}
\label{def2}
\end{equation}

Hence, by virtue of Eq.~\eqref{def2}$_3$, the scalar field of rotations is expressed by the following integration formula

\begin{equation}
\varphi\di\sss
=\varphi\di{0}
+
\integrale{0}{\sss}
\chi_\Bw\di\xi
\,\der\xi\,
\label{fm:rot}
\end{equation}

Let us consider the subsequent additive decomposition where Eq.~\eqref{def2}$_{1, 2}\,$ is taken into account
\begin{equation}
\Bu'
=(\Bu'\punto\t)\,\t
+(\Bu'\punto\tort)\,\tort
=\ee_\Bw\,\t
+(\gamma_{\cinematica} + \varphi)\,\tort\,
\label{fm:decomp}
\end{equation}
with $\,\varphi\,$ given by Eq.~\eqref{fm:rot}.

By integrating the previous equation, the displacements field $\,\Bu\,$ can be expressed as follows
\begin{equation}
\begin{split}
\Bu\di\sss
&=\Bu\di{0}
+\integrale{0}{\sss}\Bu'\der\xi\,		\\
&=\Bu\di{0}
+
\integrale{0}{\sss}
[\ee_\Bw\di\xi\,\t\di\xi
+ (\gamma_{\cinematica}\di\xi + \varphi\di\xi) \,\tort\di\xi]
\,\der\xi\,
\end{split}
\label{fm:spost}
\end{equation}

Thus, the local-nonlocal mixture elastostatic problem of a Timoshenko curved beam is composed of the differential equilibrium equations~\eqref{fm:DiffProbl} equipped with the boundary conditions~\eqref{fm:StatBC}, the constitutive equations provided by the integral convolutions in Eq.~\eqref{StressConv12} (or by the equivalent differential problem in Eqs.~\eqref{EqDiff}-\eqref{BCS}) and the kinematic compatibility condition expressed by Eqs.~\eqref{fm:rot} and~\eqref{fm:spost}. 
The integration constants $\,\varphi\di{0}\in\Re\,$ and $\,\Bu\di{0}\in\VV\,$ are univocally evaluated by prescribing essential kinematic boundary conditions.

\section{Case-studies}
\label{casi}

The procedure to solve the elastostatic problem of a Timoshenko nonlocal curved beam is now illustrated with
reference to some cases of applicative interest in Nanotechnology. 

A silicon carbide nanobeam, with Euler-Young modulus $E = 427\:$ [GPa] and
Poisson ratio $\nu = 0.2$ is considered. The beam axis is assumed to be a circle arc 
of radius $r = 10 \:$ [nm] so that the beam length is $L = r\,\pi/2$. The cross-section is a rectangular domain
$\Omega$ of base $b = 5~$[nm] and height $h = 2L/3$. In the present section, the following non-dimensional nonlocal parameter is adopted

\begin{equation}
\lambda = \frac{l_c}{L}
\label{lam}
\end{equation}
where $l_c$ is the nonlocal characteristic length parameter. 

A simple procedure is proposed in order to solve the elastostatic problem of the curved nanobeam introduced above and to investigate  size-dependent responses. It consists of the following steps.
\begin{itemize}

\item
\textbf{Step 1.}
Solution of the differential equilibrium problem in Eqs.~\eqref{fm:DiffProbl}-\eqref{fm:StatBC}
to obtain the stress fields composed of axial force $N$, shear force $T$ and bending moment $M$ as functions of $n$ integration constants, with $n$ standing for redundancy degree.
For statically determinate beams for which $n = 0$, the stress fields are univocally determined by equilibrium requirements.

\item
\textbf{Step 2.}
Evaluation of nonlocal strain fields by
applying the nonlocal stress-driven mixture model in Eq.~\eqref{StressConv12}
or by solving the equivalent differential problem governed by Eqs.~\eqref{EqDiff}-\eqref{BCS}.

\item
\textbf{Step 3.}
Detection of curved nanobeam nonlocal displacements $\Bu$ and rotations $\varphi$
fields by imposing $3 + n$ essential kinematic boundary conditions.

 \end{itemize}

Note that in the case-studies presented in this section, total
deformations and elastic deformation fields are assumed to be coincident.


\medskip
\medskip
\medskip

\noindent\textbf{Cantilever beam under point-force at free end}

\begin{figure}[h]
\centering	
\includegraphics[width=0.55\textwidth]{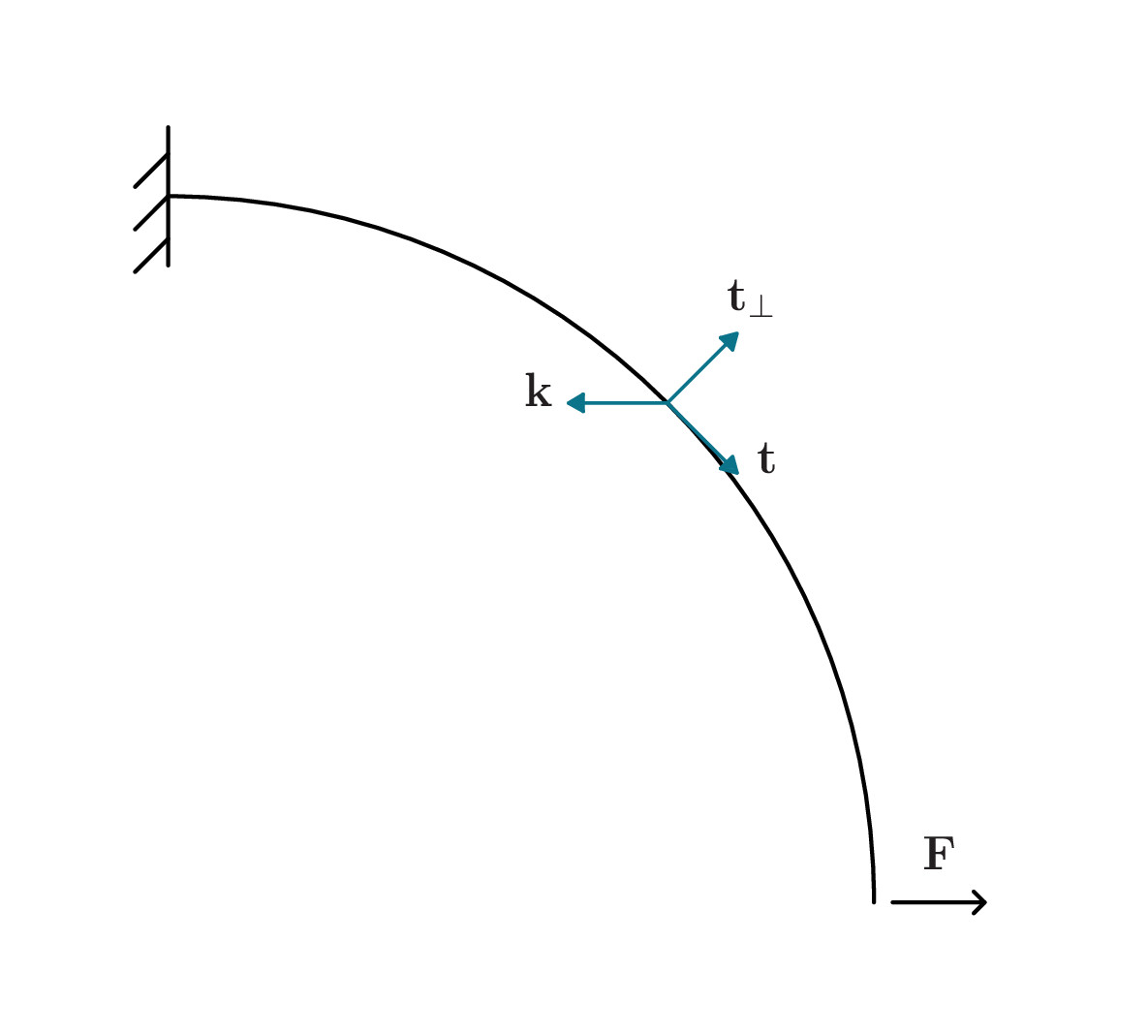}
\caption{Cantilever curved nanobeam.}
\label{schema1}
\end{figure}

\noindent The curved nanobeam introduced above is clamped at the abscissa $\, s = 0\,$ and subjected to a force $\,F = 10\,$ [nN]
applied at free end, as shown in Fig.~\ref{schema1}. Hence, the essential kinematic boundary conditions are

\begin{equation}
\Bu\di{0}=\Bo\, \qquad
\varphi\di{0}=0\,
\label{kBC}
\end{equation}
From Eqs.~\eqref{fm:StatBC} and \eqref{kBC} follows that natural static boundary conditions are expressed by
\begin{equation}
\begin{cases}
N(L)&= 0\,\\
T(L)&= F\, \\
M(L)&= 0\,
\label{s.c.}
\end{cases}
\end{equation}

\begin{figure}[!ht]
\centering
\subfloat[][\emph{Transversal displacement $u_{\tort}$ as fuction of $\lambda$ for $m = 0.3$.}]
{\includegraphics[width=.5\textwidth]{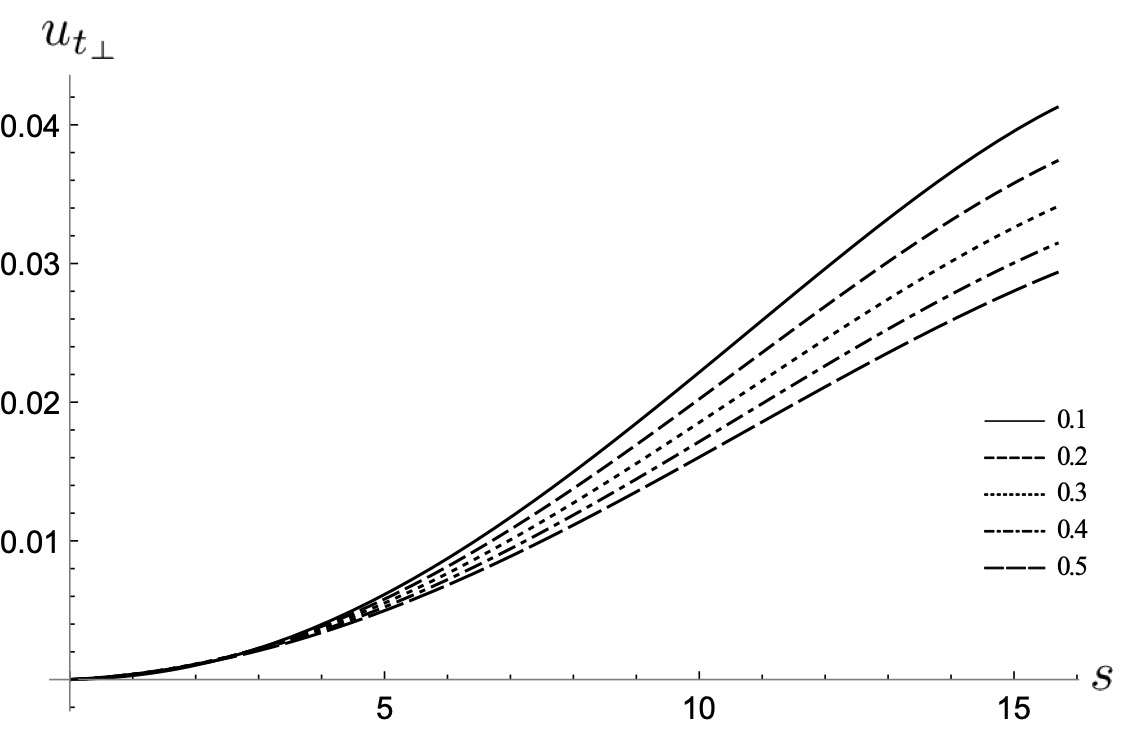}} \quad
\subfloat[][\emph{Transversal displacement $u_{\tort}$ as fuction of $\lambda$ for $m = 0.6$.}]
{\includegraphics[width=.5\textwidth]{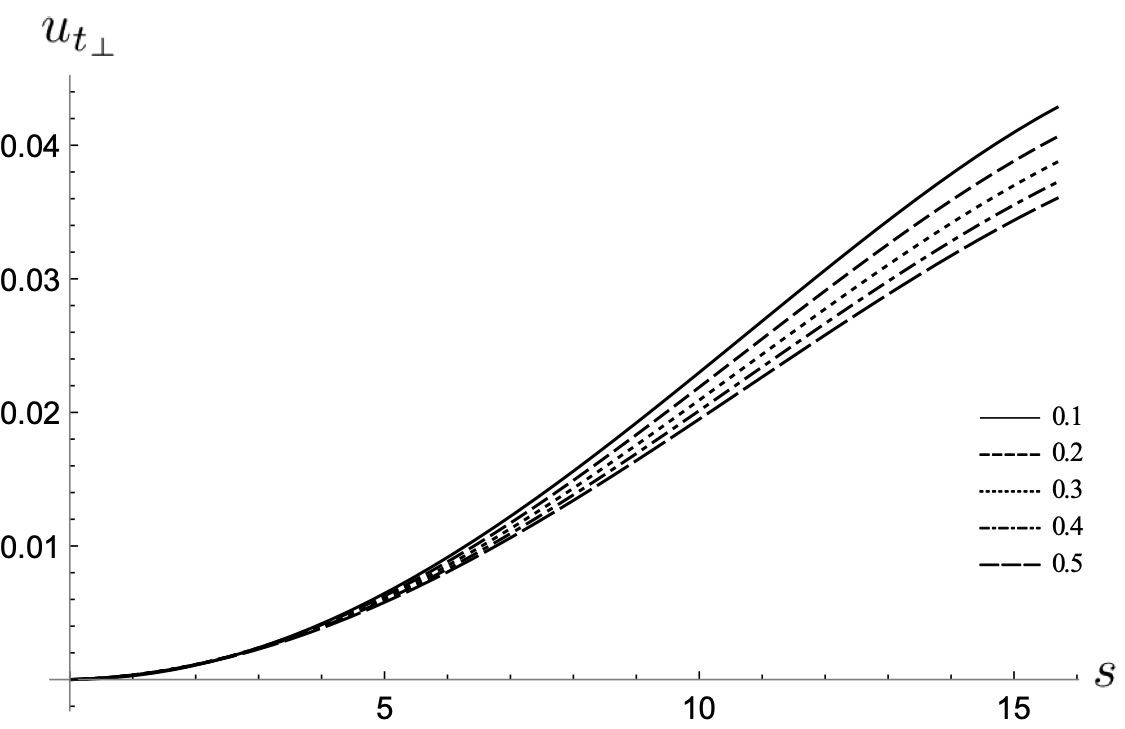}} 
\caption{Cantilever beam: nonlocal transversal displacement $u_{\tort}$ [nm] versus $s \in [0, L]$ [nm] as fuction of $\lambda$.}
\label{d1}
\end{figure}

\begin{figure}[!h]
\centering
\subfloat[][\emph{Axial displacement $u_{\t}$ as fuction of $\lambda$ for $m = 0.3$.}]
{\includegraphics[width=.5\textwidth]{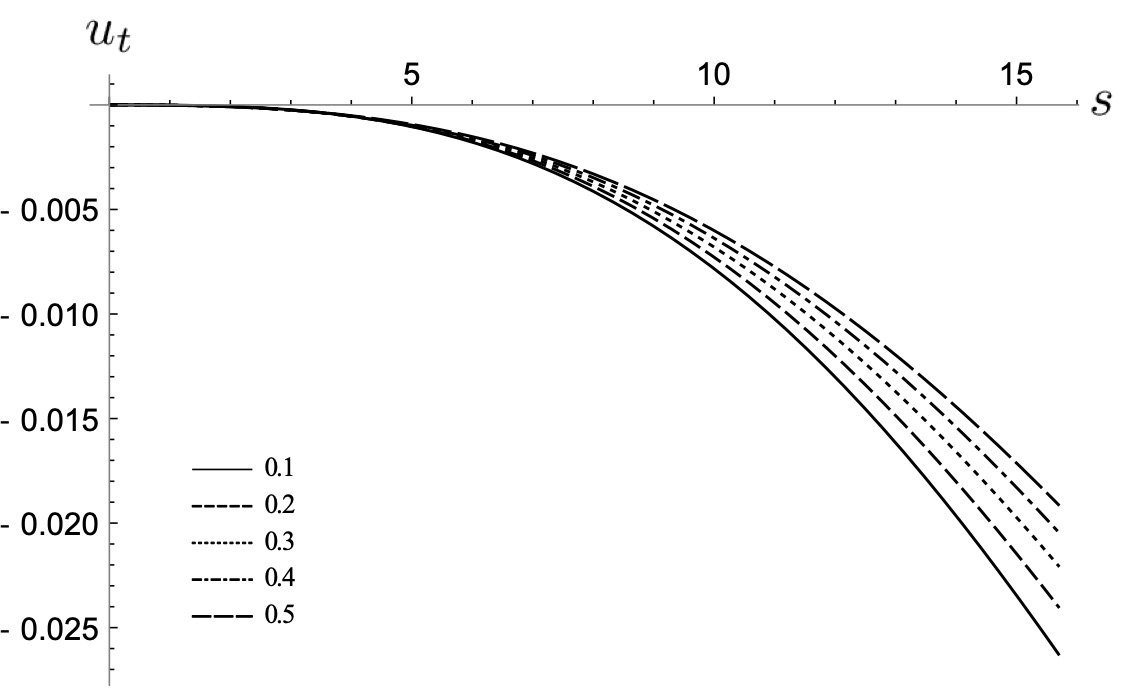}} \quad
\subfloat[][\emph{Axial displacement $u_{\t}$ as fuction of $\lambda$ for $m = 0.6$.}]
{\includegraphics[width=.5\textwidth]{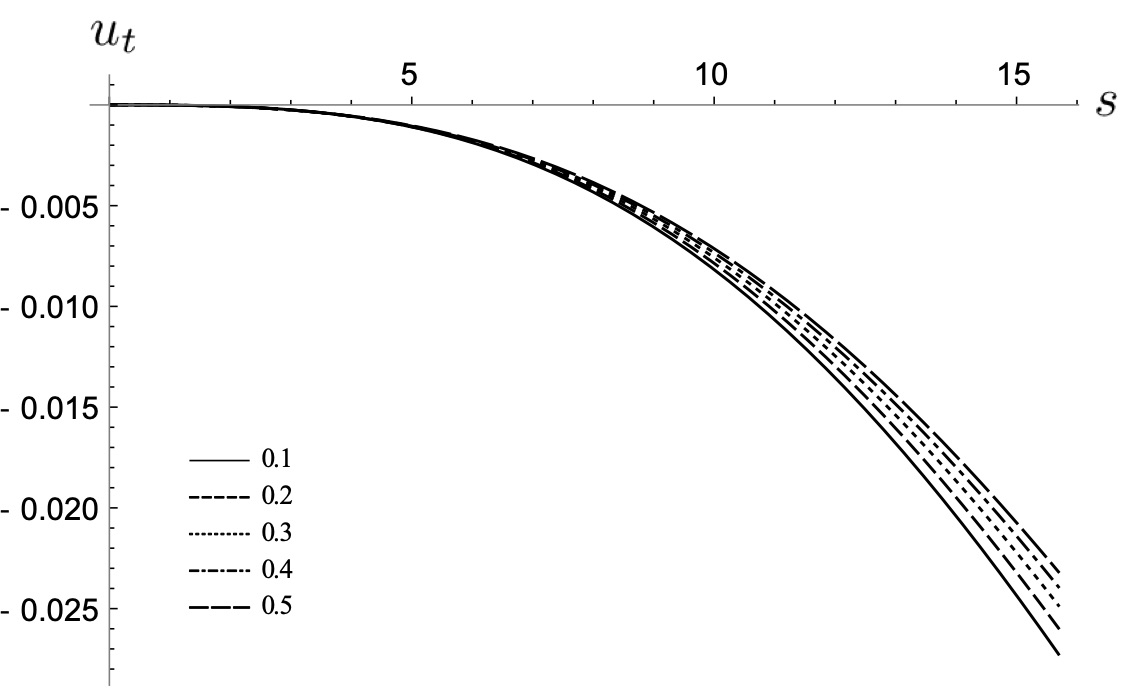}} 
\caption{Cantilever beam: nonlocal axial displacement $u_{\t}$  [nm] versus $s \in [0, L]$  [nm] as fuction of $\lambda$.}
\label{d2}
\end{figure}

\begin{figure}[!h]
\centering
\subfloat[][\emph{Bending rotation $\varphi$ as fuction of $\lambda$ for $m = 0.3$.}]
{\includegraphics[width=.5\textwidth]{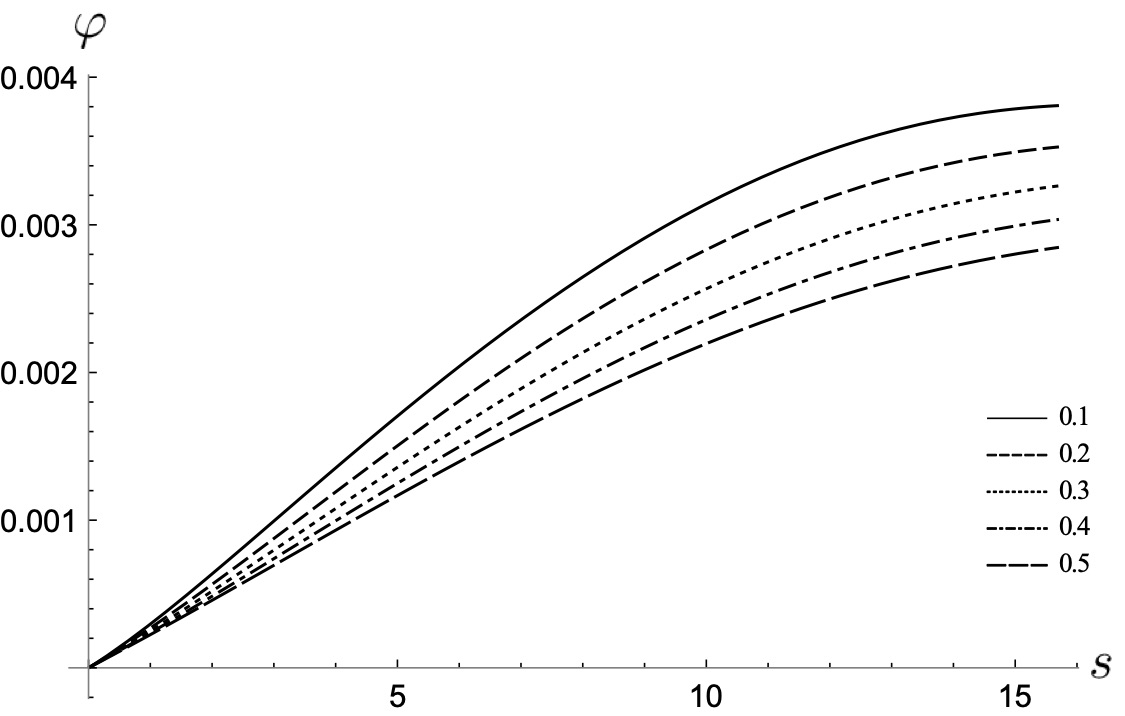}} \quad
\subfloat[][\emph{Bending rotation $\varphi$ as fuction of $\lambda$ for $m = 0.6$.}]
{\includegraphics[width=.5\textwidth]{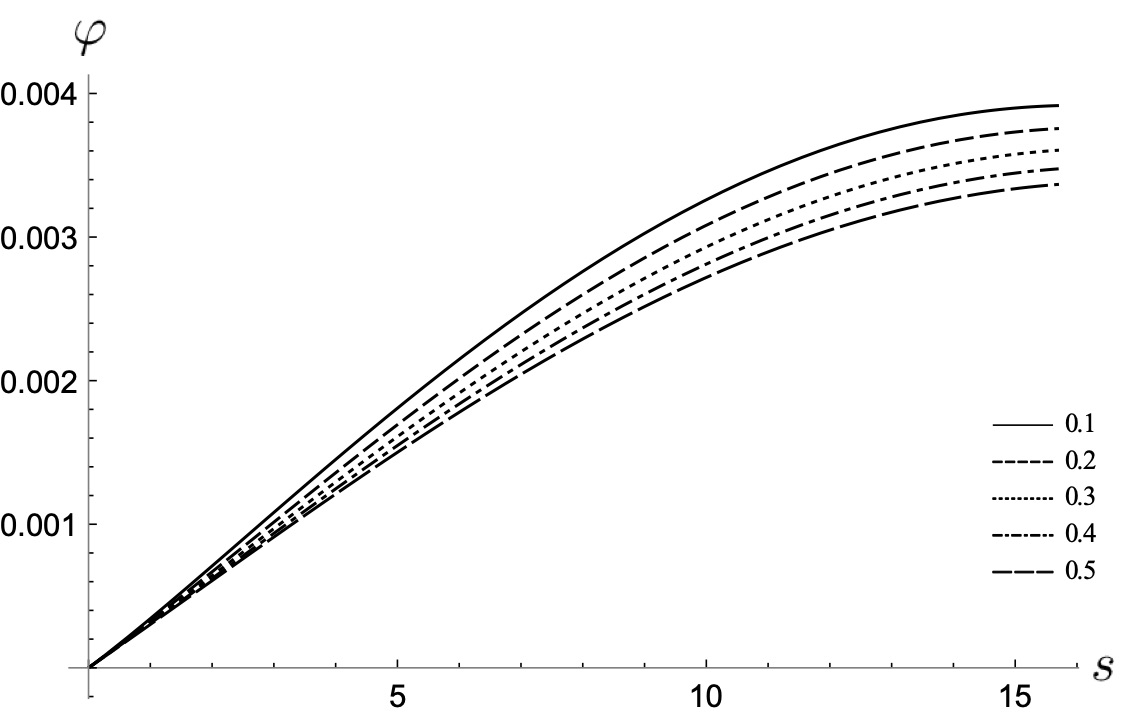}} 
\caption{Cantilever beam: nonlocal bending rotation $\varphi$ [-] versus $s \in [0, L]$  [nm] as fuction of $\lambda$.}
\label{d3}
\end{figure}

Since the redundancy degree $\,n\,$ is vanishing, by following $Step~1$ the stress fields of axial force $N$, shear force $T$ and bending moment $M$ are
univocally determined by equilibrium requirements in Eqs.~\eqref{fm:DiffProbl}-\eqref{fm:StatBC} and their parametric expressions are provided below:
\begin{equation}
\begin{cases}
&N(s) = F\,\cos \big{(}s/r\big{)} \\
&T(s) = F\,\sin \big{(}s/r\big{)} \\
&M(s) = F\,r\cos \big{(}s/r\big{)}
\end{cases}
\label{formuleParam}
\end{equation}
Transversal displacements $u_{t_{\bot}} = \Bu \cdot \tort$, axial displacements $u_t =\Bu \cdot \t$ and bending rotations $\varphi$ are obtained by following $Steps~2$-$\,3$ and then plotted
as fuction of $s \in [0, L]$ as shown in Figs.~\ref{d1} - \ref{d3}. In particular Figs.~\ref{d1} - \ref{d3} represent displacements and rotations of the nanocantilever for increasing values of the nonlocal parameter $\lambda$, for fixed values of the mixture parameter. As also shown by  numerical results in Tab.~\ref{tab1}, the response stiffens for increasing values of the nonlocal parameter $\lambda$ and 
exhibits a softening behavior by increasing the mixture parameter $m$.

\medskip
\medskip

\noindent\textbf{Slider and roller supported beam under uniformly distributed loading}

\begin{figure}[h]
\centering	
\includegraphics[width=0.55\textwidth]{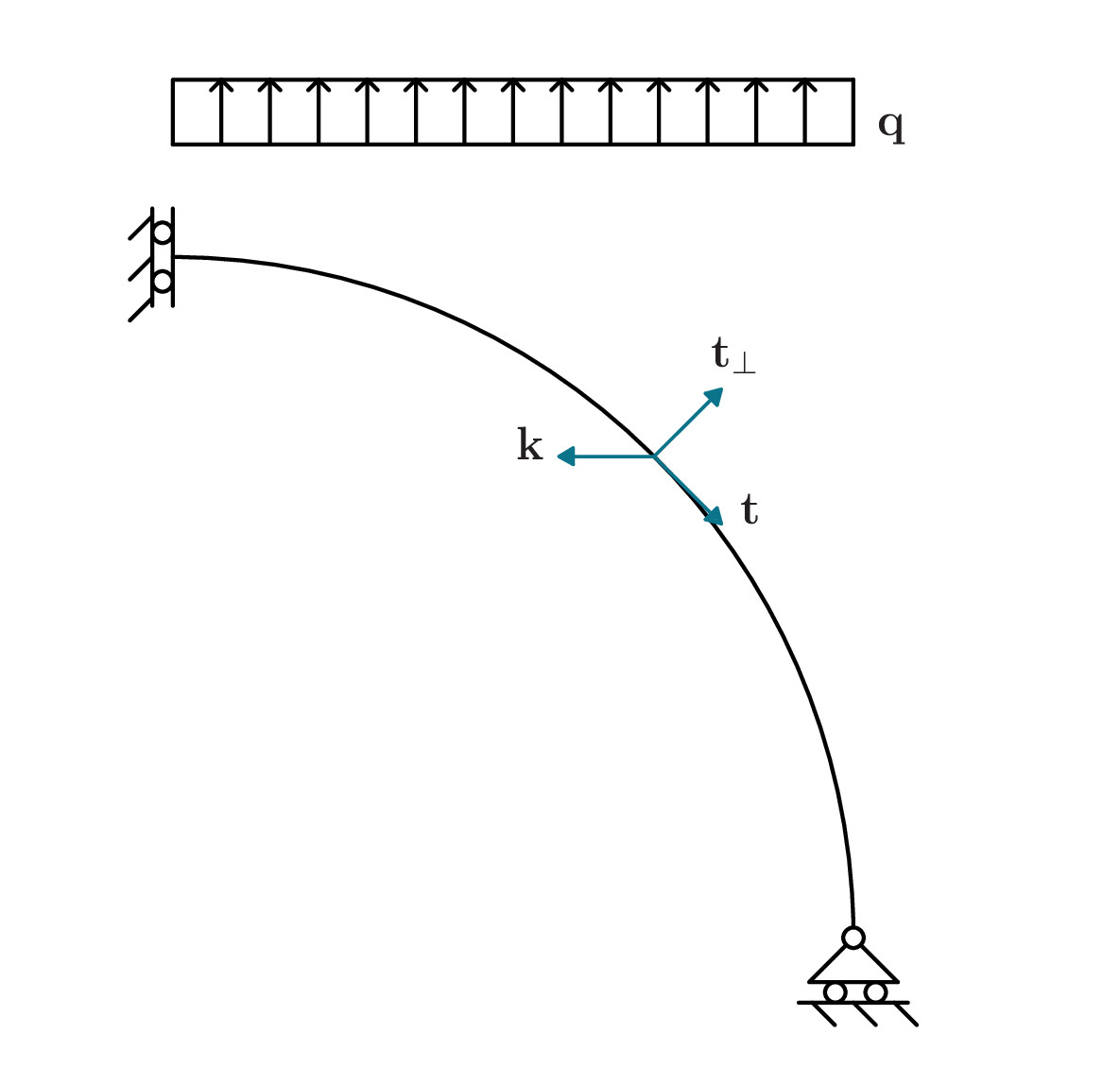}
\caption{Curved nanobeam with slider and roller supports.}
\label{schema2}
\end{figure}

\noindent The curved nanobeam has a slider and a roller supports, at $s=0$ and $s=L$ 
respectively. It is subjected to a uniformly distributed vertical loading $\,q = 2\,$ [nN/nm] directed upwards as shown in Fig.~\ref{schema2}. Now, the essential kinematic boundary conditions are
\begin{equation}
\Bu\di{0}\punto\t\di{0}=0\, \qquad
\varphi\di{0}=0\, \qquad
\Bu\di{L}\punto\t\di{L}=0\,
\end{equation}
From Eqs.~\eqref{fm:StatBC} and \eqref{kBC} follows that natural static boundary conditions are expressed by
\begin{equation}
\left\{\begin{split}
T(0)&= 0\,\\
T(L)&= 0\,  \\
M(L)&= 0\,
\end{split}\right.
\end{equation}

Since the beam is statically determinate, the stress fields $\,N\,,T\,,M\,$ are
univocally obtained by equilibrium conditions and their parametric expressions as function of the arch length are shown as follows:

\begin{equation}
\begin{cases}
&N(s) = q\, r \sin ^2 \big{(}s/r\big{)} \\
&T(s) = -\frac{1}{2} q\, r \sin \big{(}2s/r\big{)}  \\
&M(s) = -\frac{1}{2} q\, r^2 \cos ^2 \big{(}s/r\big{)}
\end{cases}
\label{formuleParam2}
\end{equation}

\begin{figure}[!h]
\centering
\subfloat[][\emph{Transversal displacement $u_{\tort}$ as fuction of $\lambda$ for $m = 0.3$.}]
{\includegraphics[width=.5\textwidth]{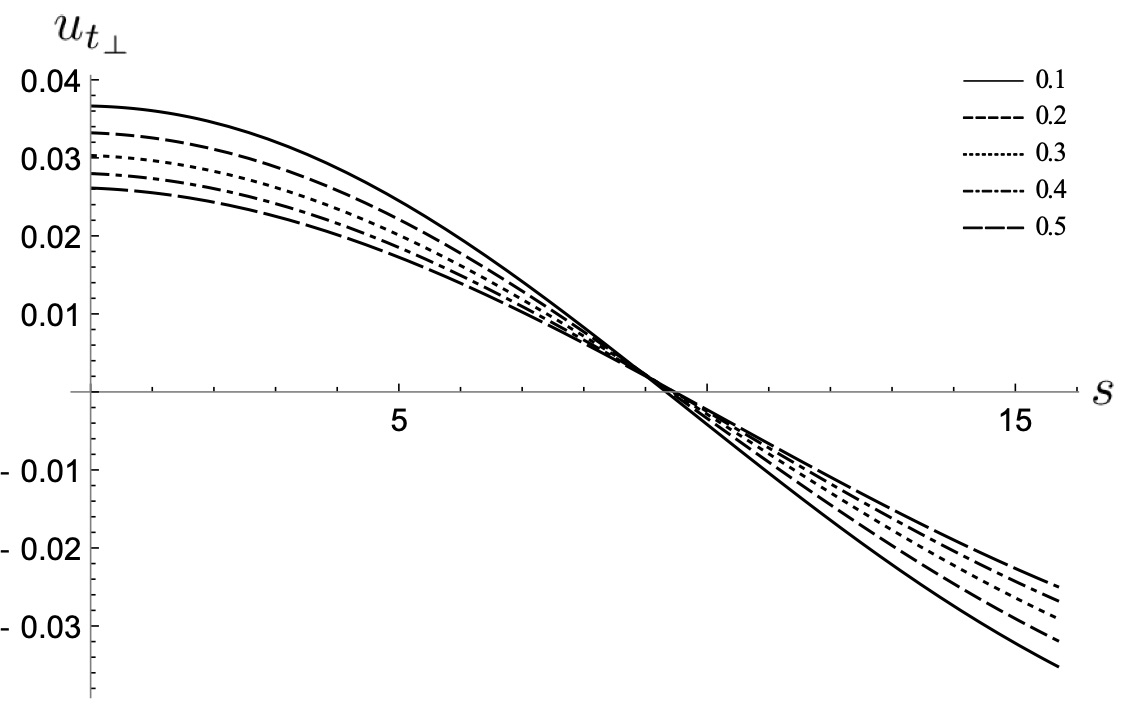}} \quad
\subfloat[][\emph{Transversal displacement $u_{\tort}$ as fuction of $\lambda$ for $m = 0.6$.}]
{\includegraphics[width=.5\textwidth]{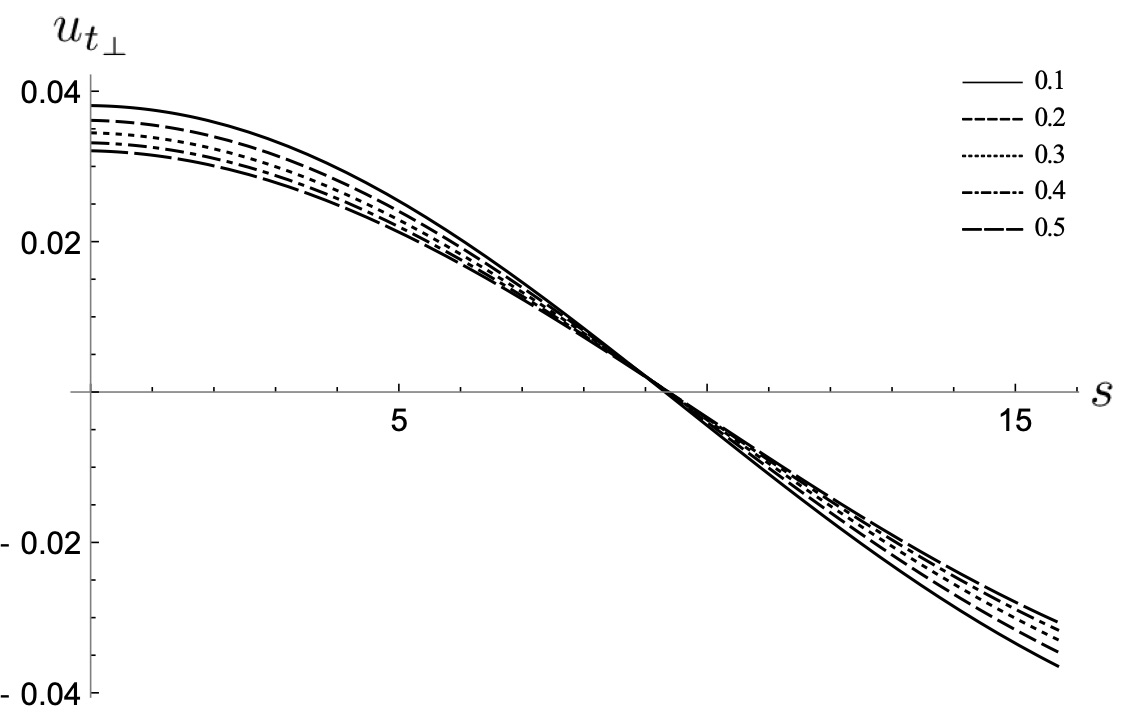}} 
\caption{Slider and roller supported beam: nonlocal transversal displacement $u_{\tort}$ [nm] versus $s \in [0, L]$ [nm] as fuction of $\lambda$.}
\label{d12}
\end{figure}

\begin{figure}[!h]
\centering
\subfloat[][\emph{Axial displacement $u_{\t}$ as fuction of $\lambda$ for $m = 0.3$.}]
{\includegraphics[width=.5\textwidth]{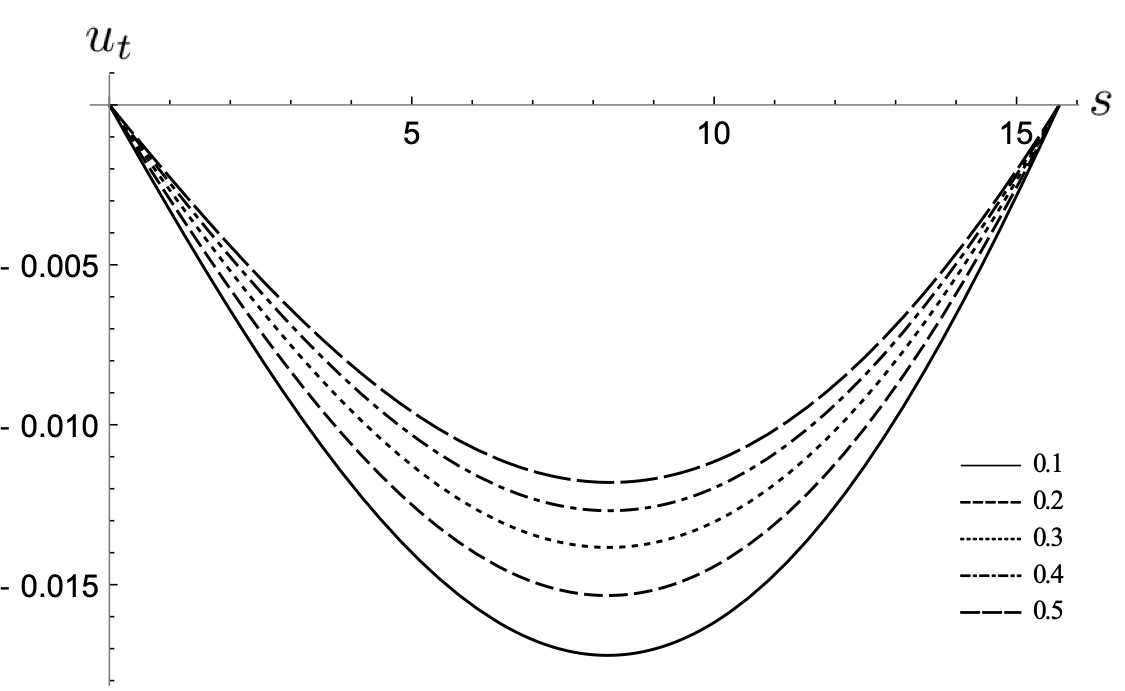}} \quad
\subfloat[][\emph{Axial displacement $u_{\t}$ as fuction of $\lambda$ for $m = 0.6$.}]
{\includegraphics[width=.5\textwidth]{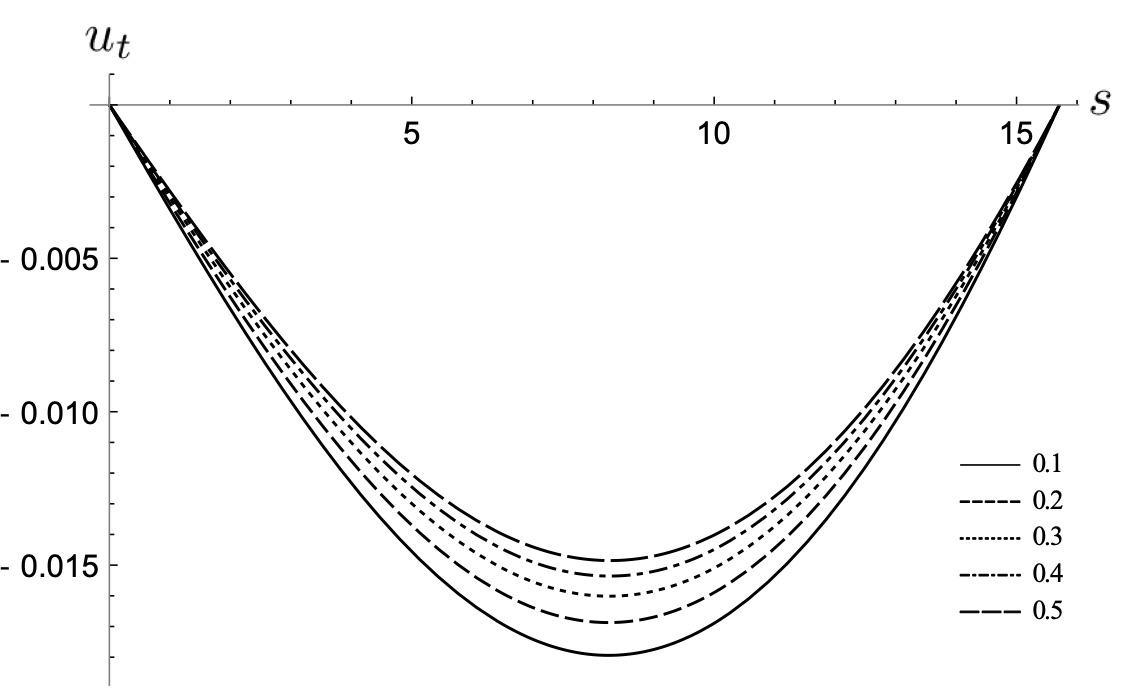}} 
\caption{Slider and roller supported beam: nonlocal axial displacement $u_{\t}$  [nm] versus $s \in [0, L]$  [nm] as fuction of $\lambda$.}
\label{d22}
\end{figure}

\begin{figure}[!h]
\centering
\subfloat[][\emph{Bending rotation $\varphi$ as fuction of $\lambda$ for $m = 0.3$.}]
{\includegraphics[width=.5\textwidth]{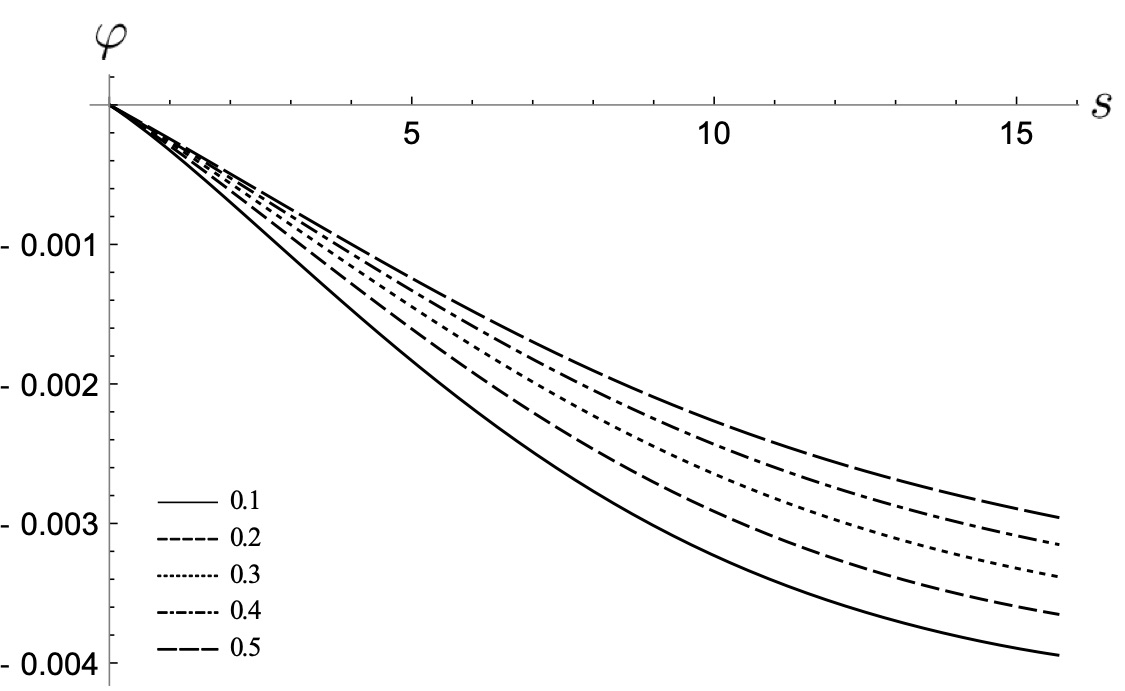}} \quad
\subfloat[][\emph{Bending rotation $\varphi$ as fuction of $\lambda$ for $m = 0.6$.}]
{\includegraphics[width=.5\textwidth]{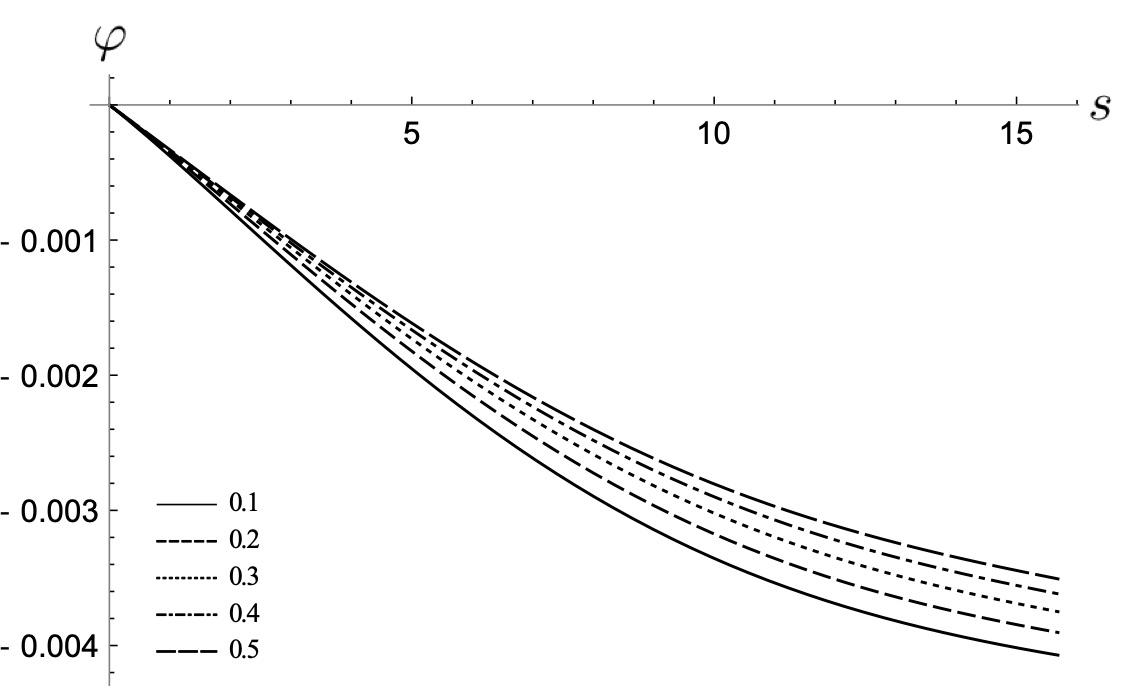}} 
\caption{Slider and roller supported beam: nonlocal bending rotation $\varphi$ [-] versus $s \in [0, L]$  [nm] as fuction of $\lambda$.}
\label{d32}
\end{figure}

Then, the transversal displacements $u_{t_{\bot}} = \Bu \cdot \tort$, axial displacements $u_t =\Bu \cdot \t$ and bending rotations $\varphi$ are obtained by following $Steps~2$-$\,3$ and represented
in Figs.~\ref{d12} - \ref{d32} as function of $\lambda$ for fixed values of the mixture parameters $m$. Numerical results in terms of displacements and rotations are provided in Tab.~\ref{tab2}.
\medskip
\medskip

\noindent \textbf{Clamped and roller supported beam under uniformly distributed loading}

\begin{figure}[h]
\centering	
\includegraphics[width=0.55\textwidth]{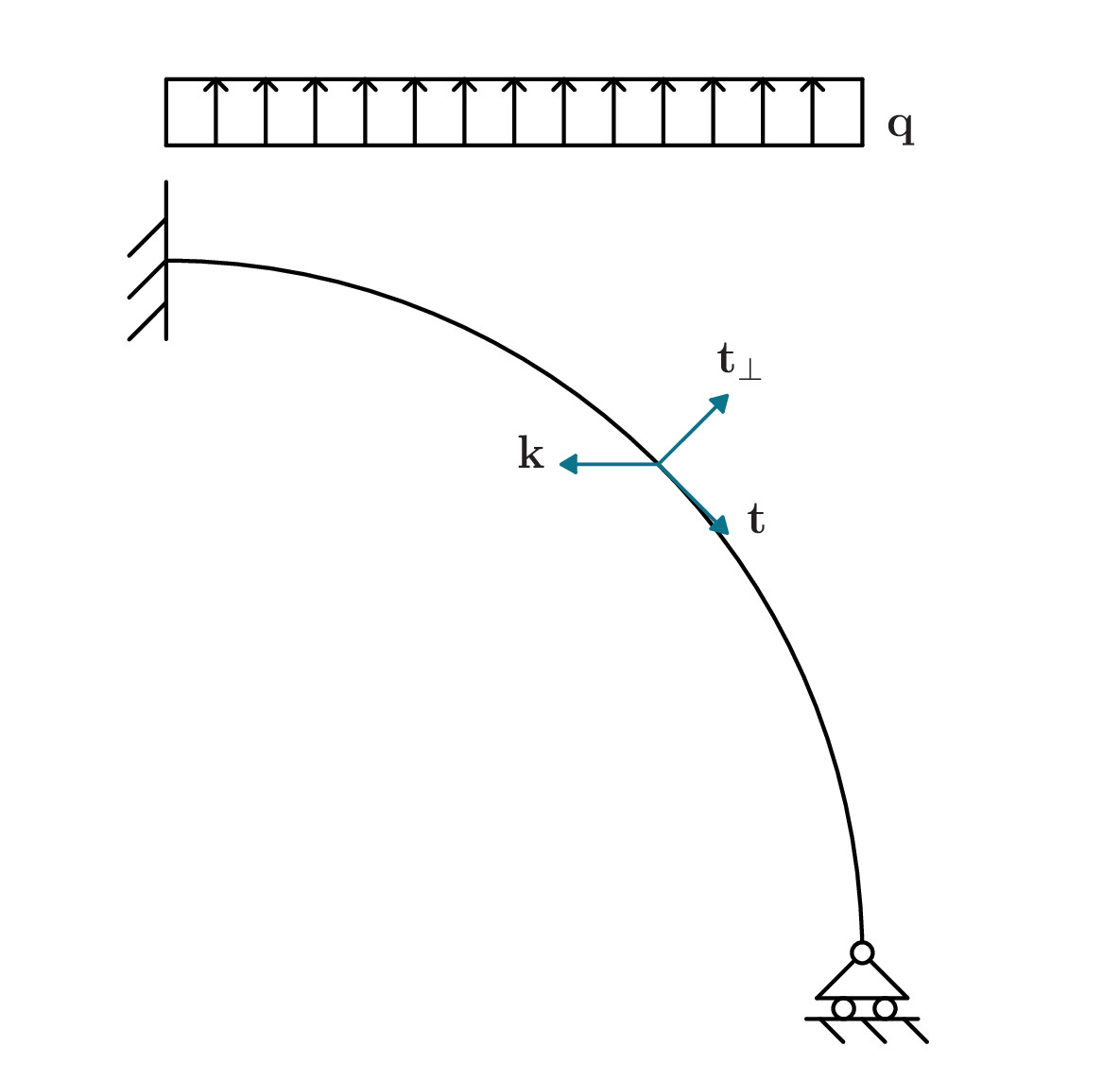}
\caption{Clamped and roller supported curved nanobeam.}
\label{schema3}
\end{figure}

\noindent Let us consider a curved nanobeam with clamped and roller supported ends, at $s=0$ and $s=L$ 
respectively. It is subjected to a uniformly distributed vertical loading $\,q = 5\,$ [nN/nm] directed upwards as shown in Fig.~\ref{schema3}. Hence, the essential kinematic boundary conditions are
\begin{equation}
\Bu\di{0}=0\, \qquad
\varphi\di{0}=0\, \qquad
\Bu\di{L}\punto\t\di{L}=0\,
\label{kBC3}
\end{equation}
From Eqs.~\eqref{fm:StatBC} and \eqref{kBC3} the corrisponding natural static boundary conditions take the form
\begin{equation}
\left\{\begin{split}
T(L)&= 0\,\\
M(L)&= 0\,  
\end{split}\right.
\label{sBC3}
\end{equation}

\begin{figure}[h!]
\centering	
\includegraphics[width=0.5\textwidth]{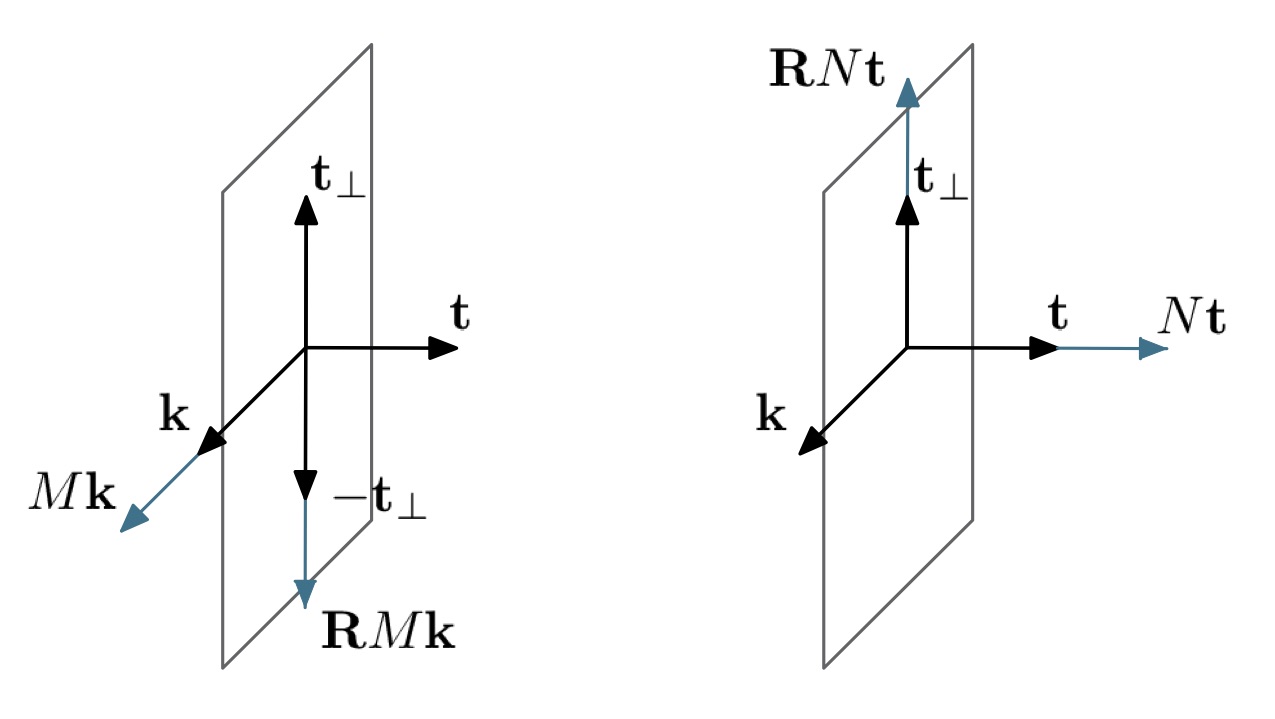}
\caption{Sketch of cross section with local triad and bending $\R M\k\,$ and axial $\R N\t$ vectors.}
\label{ske2}
\end{figure}

\begin{figure}[!h]
\centering	
\includegraphics[width=0.51\textwidth]{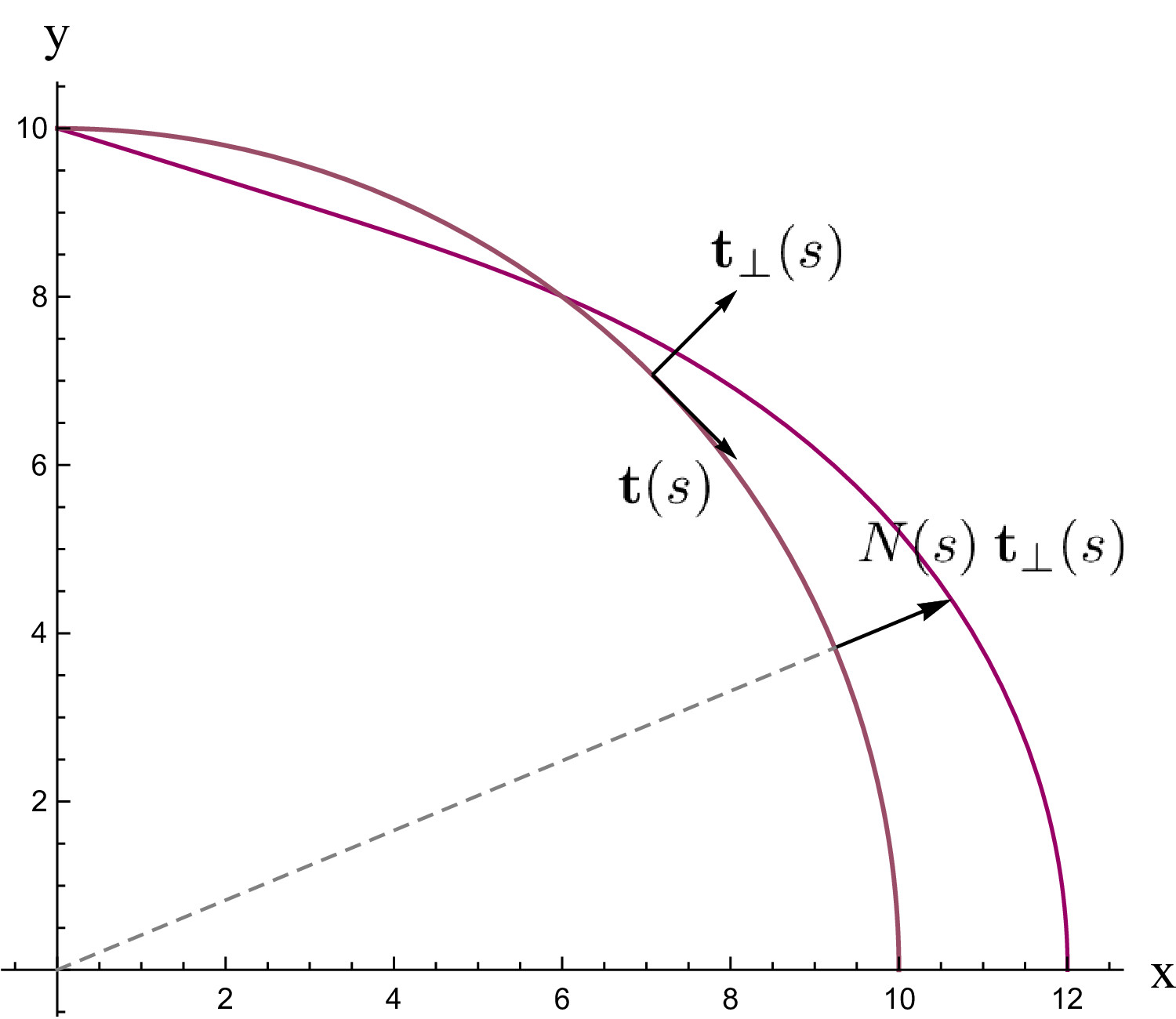}
\caption{Clamped and roller supported beam: plot of the vector field $\R N\t = N\tort$ [10\,nN] in the Cartesian plane x [nm], y [nm] for $\lambda = 0.5$ and $m = 0.3$.}
\label{N03}
\end{figure}

\begin{figure}[!h]
\centering	
\includegraphics[width=0.48\textwidth]{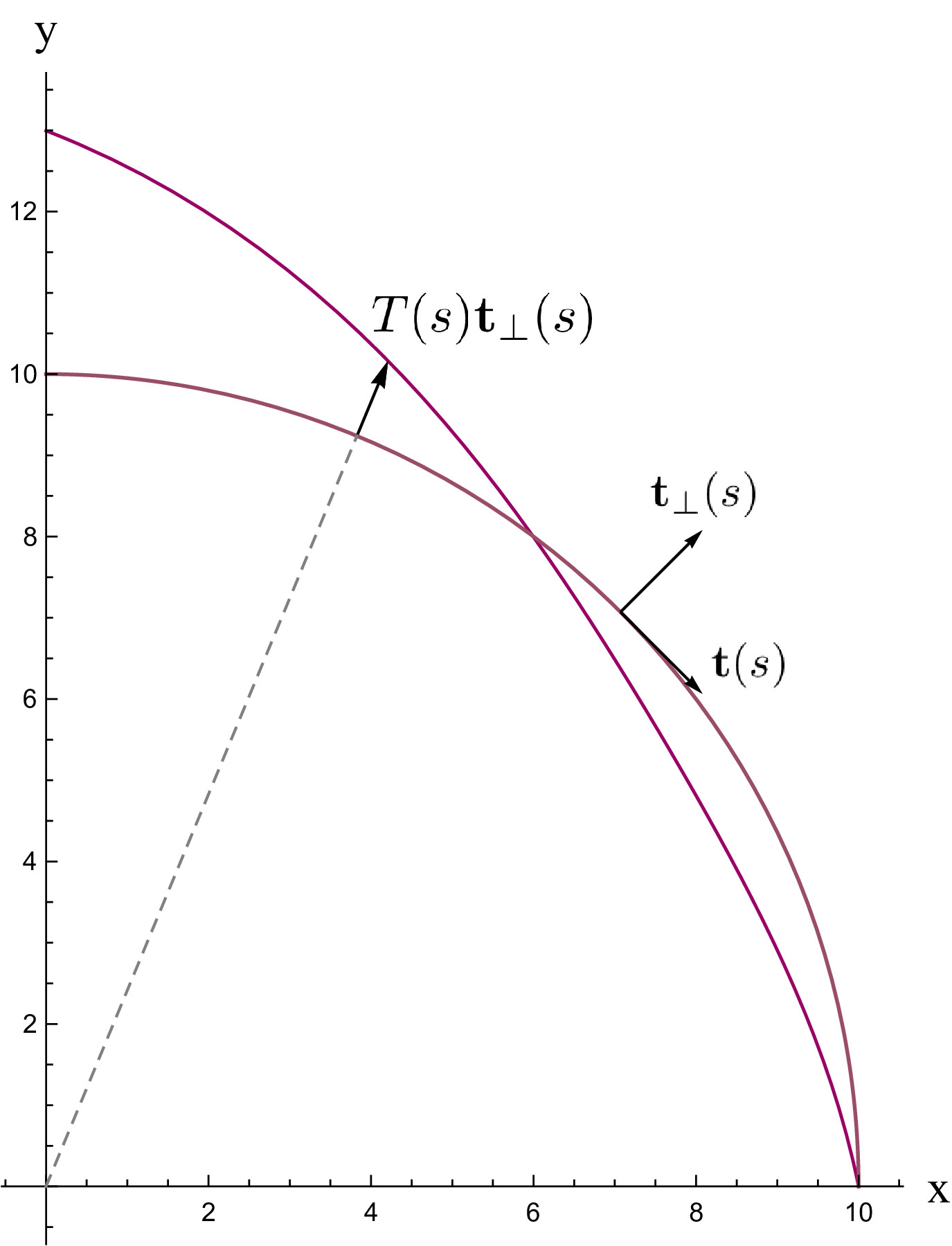}
\caption{Clamped and roller supported beam: plot of the vector field $T\tort$ [10\,nN] in the Cartesian plane x [nm], y [nm] for $\lambda = 0.5$ and $m = 0.3$.}
\label{T03}
\end{figure}

\begin{figure}[!h]
\centering	
\includegraphics[width=0.5\textwidth]{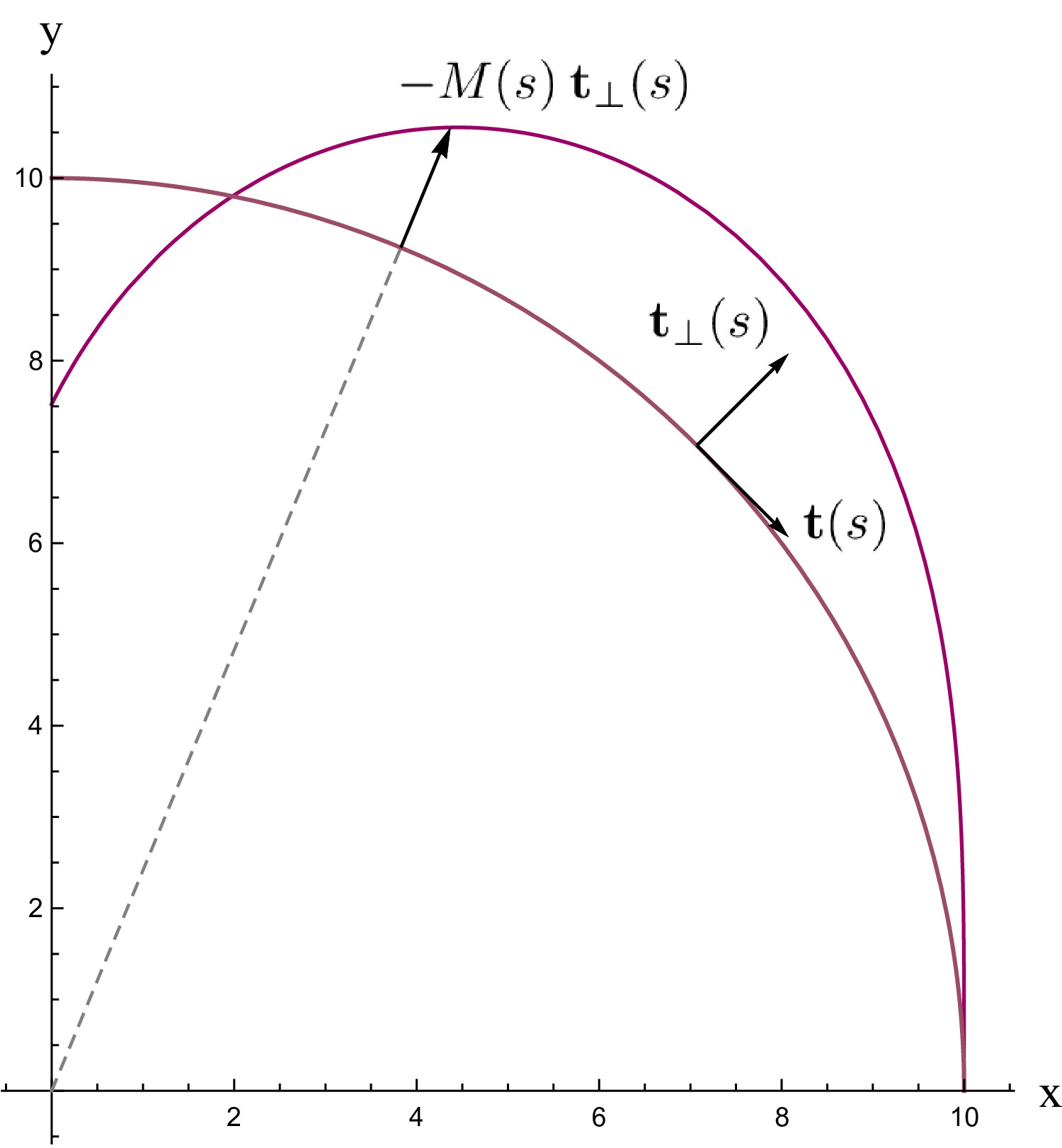}
\caption{Clamped and roller supported beam: plot of the vector field $\R M\k = -M\tort$ [20\,nN\,nm] in the Cartesian plane x [nm], y [nm] for $\lambda = 0.5$ and $m = 0.3$.}
\label{M03}
\end{figure}

\begin{figure}[!h]
\centering
\subfloat[][\emph{Transversal displacement $u_{\tort}$ as fuction of $\lambda$ for $m = 0.3$.}]
{\includegraphics[width=.5\textwidth]{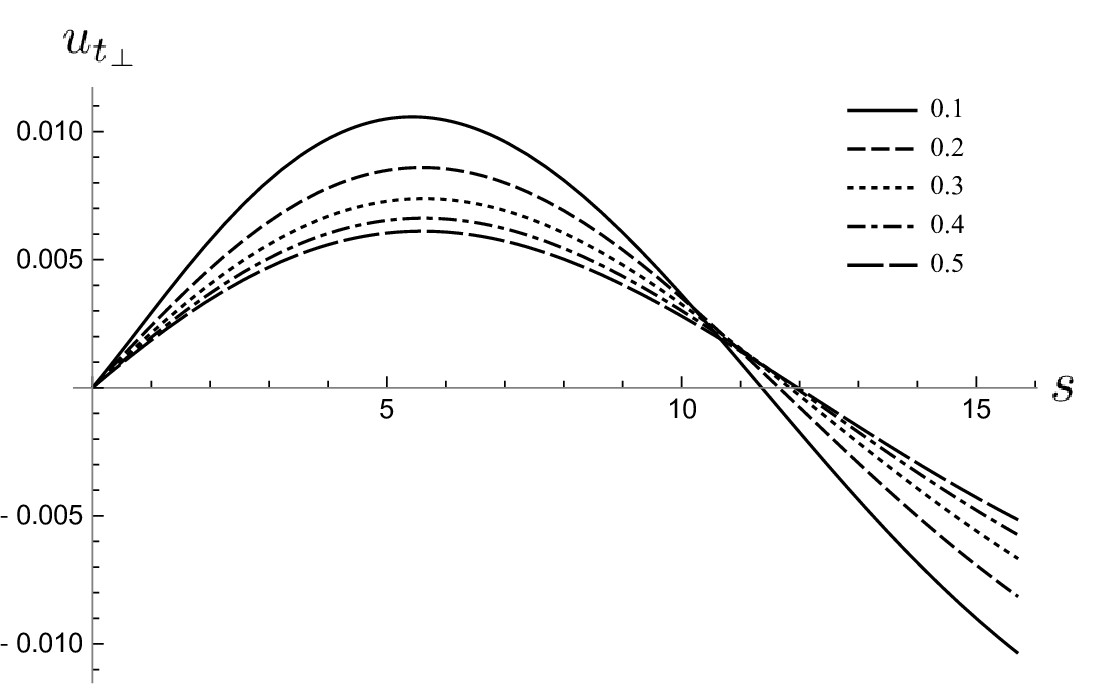}} \quad
\subfloat[][\emph{Transversal displacement $u_{\tort}$ as fuction of $\lambda$ for $m = 0.6$.}]
{\includegraphics[width=.5\textwidth]{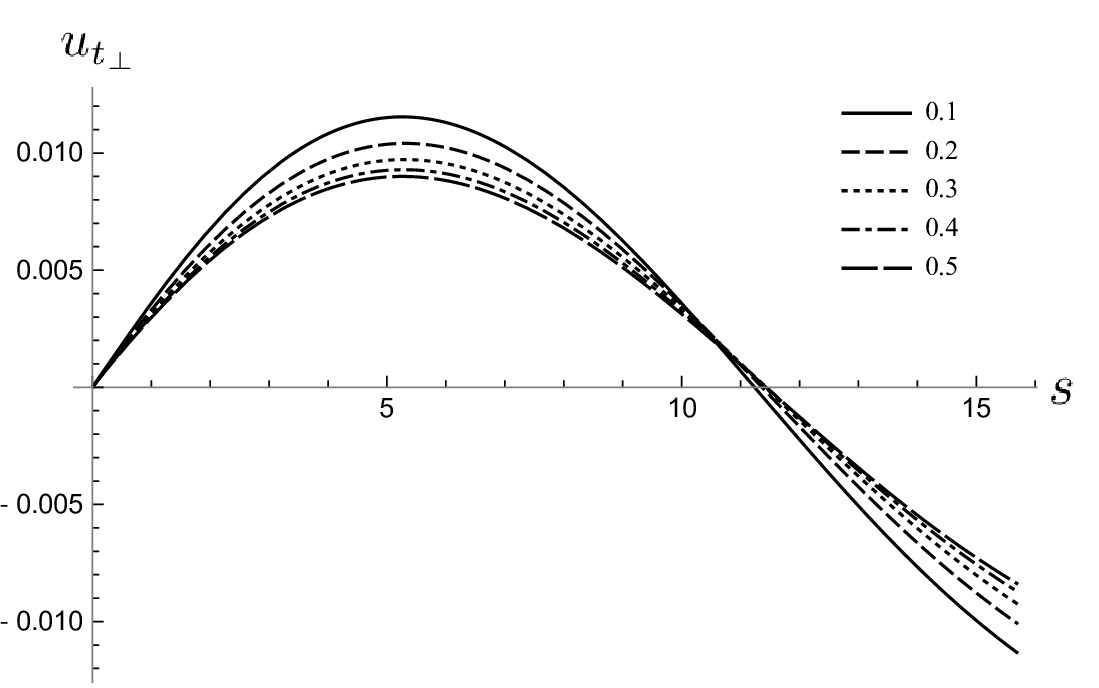}} 
\caption{Clamped and roller supported beam: nonlocal transversal displacement $u_{\tort}$ [nm] versus $s \in [0, L]$ [nm] as fuction of $\lambda$.}
\label{3d1}
\end{figure}

\begin{figure}[!h]
\centering
\subfloat[][\emph{Axial displacement $u_{\t}$ as fuction of $\lambda$ for $m = 0.3$.}]
{\includegraphics[width=.5\textwidth]{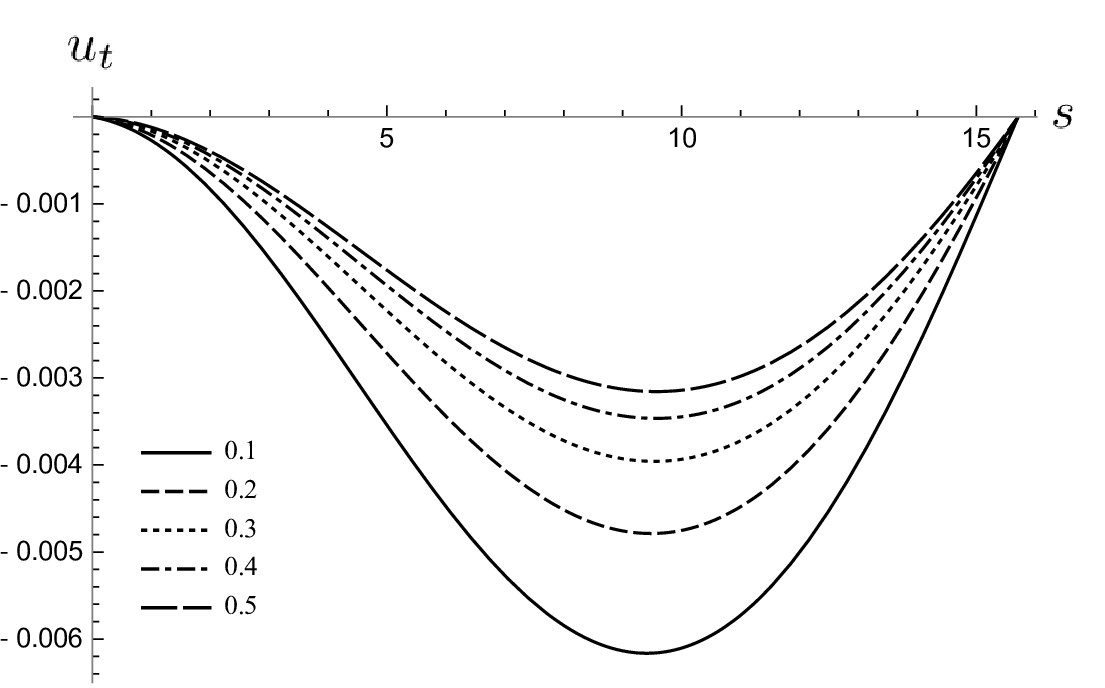}} \quad
\subfloat[][\emph{Axial displacement $u_{\t}$ as fuction of $\lambda$ for $m = 0.6$.}]
{\includegraphics[width=.5\textwidth]{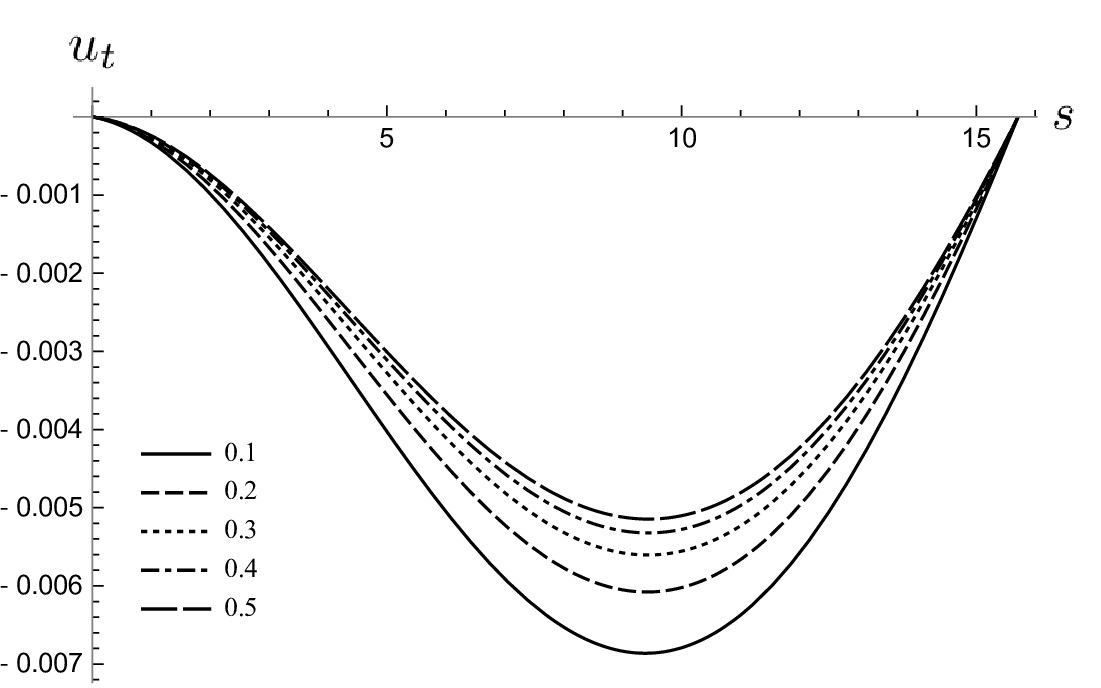}} 
\caption{Clamped and roller supported beam: nonlocal axial displacement $u_{\t}$  [nm] versus $s \in [0, L]$  [nm] as fuction of $\lambda$.}
\label{3d2}
\end{figure}

\begin{figure}[!h]
\centering
\subfloat[][\emph{Bending rotation $\varphi$ as fuction of $\lambda$ for $m = 0.3$.}]
{\includegraphics[width=.5\textwidth]{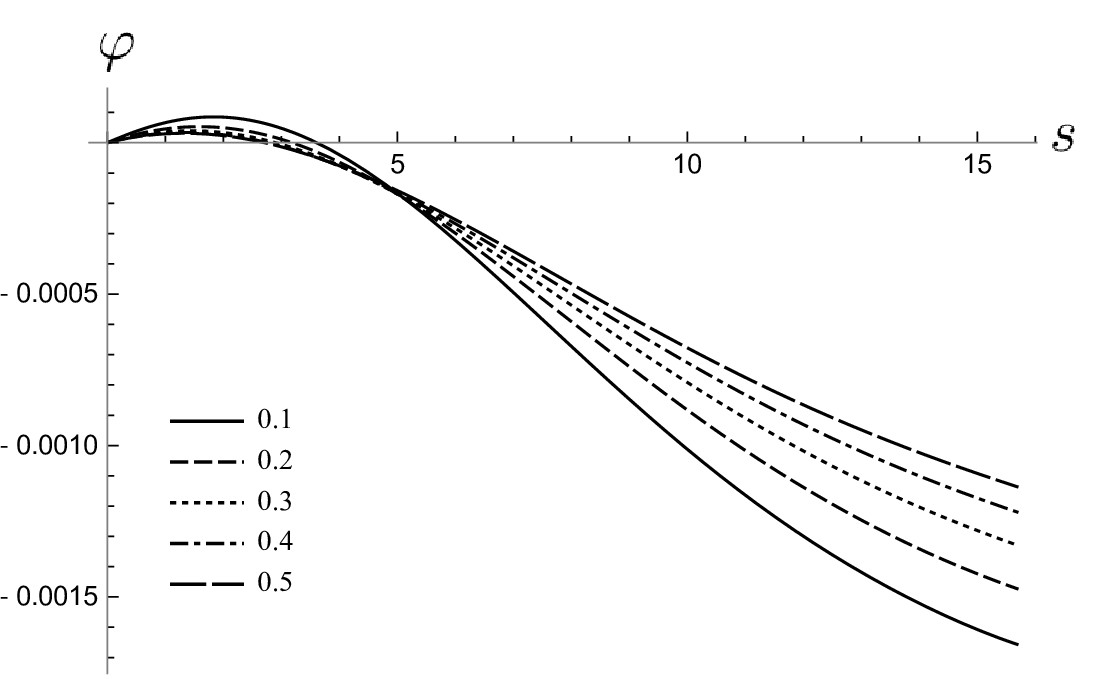}} \quad
\subfloat[][\emph{Bending rotation $\varphi$ as fuction of $\lambda$ for $m = 0.6$.}]
{\includegraphics[width=.5\textwidth]{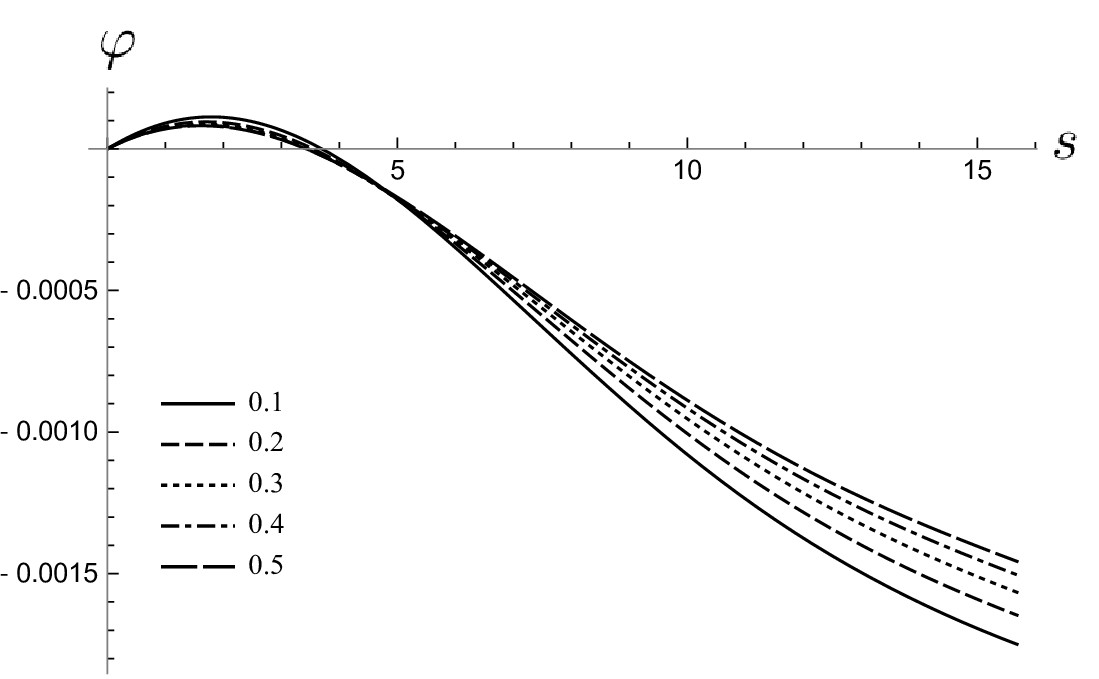}} 
\caption{Clamped and roller supported beam: nonlocal bending rotation $\varphi$ [-] versus $s \in [0, L]$  [nm] as fuction of $\lambda$.}
\label{3d3}
\end{figure}

Axial force $N$, shear force $T$ and bending moment $M$
are solutions of the differential equilibrium problem in Eqs.~\eqref{fm:DiffProbl}-\eqref{sBC3} and since the redundancy degree is $n = 1$, the stress fields are functions of one integration costant.
Parametric nonlocal strain fields are obtained by the integral stress-driven mixture model in  Eq.~\eqref{StressConv12} or by solving the equivalent differential problem in Eqs.~\eqref{EqDiff}-\eqref{BCS}.
Then, nonlocal displacements $\Bu$ and rotations $\varphi$ are detected by imposing
the $3 + n$ essential kinematic boundary conditions in Eq.~\eqref{kBC3}.
The static fields of axial force $\,\R N\t = N\tort\,$, shear force $\,T\tort\,$ and bending moment $\,\R M\k = -M\tort\,$, parameterized in terms of the arch length $s$, are
graphically represented in Figs.~\ref{N03}, \ref{T03}, \ref{M03}, respectively. A schematic sketch of cross section with the local triad and the vectors $\R M\k\,$ and $\R N\t$ is depicted in Fig.~\ref{ske2}.
Transversal displacements $u_{t_{\bot}} = \Bu \cdot \tort$, axial displacements $u_t =\Bu \cdot \t$ and bending rotations $\varphi$ are plotted versus $s \in [0, L]$ in Figs.~\ref{3d1} - \ref{3d3}.
As it is shown in the parametric plots, the response stiffens for increasing nonlocal parameter $\lambda$ and exhibits a softening behavior by increasing the mixture parameter $m$.
This trend is also confirmed by numerical results provided in Tab.~\ref{tab3}.
%

\noindent\textbf{Remark}. It is worth noting that for vanishing geometric curvature and mixture parameters, solution procedures and results of the case-studies illustrated in this section are coincident with the ones provided by Barretta et. al\citep{BarEuropean} for straight Timoshenko nanobeams. An alternative solution methodolody for straight thick nonlocal beams based on Laplace trasform has been recently proposed by Zhang et al.\citep{Zhang2020}.

%
%

%

\begin{table*}[h!]
\begin{center}
\resizebox{1\textwidth}{!}{
\begin{tabular}{c|cc|cc|cc|cc}  
\multicolumn{1}{c}{$\lambda$}&\multicolumn{2}{c}{$u_{\bot}(L)\: [10^{-2}nm] $}&\multicolumn{2}{c}{$u_{t}(L) \:[10^{-2}nm]$}&\multicolumn{2}{c}{$\varphi(L)\: [10^{-3}]$}&\multicolumn{2}{c}{$u^{tot}_{max}(L) \:[10^{-2}nm] $} \\
$ $ & $ m = 0.3 $   &  $ m = 0.6  $ &   $ m = 0.3 $   &  $ m = 0.6  $  & $ m = 0.3 $   &  $ m = 0.6  $  & $ m = 0.3 $   &  $ m = 0.6  $  \\
\hline
\cline{1-9}
0,1	&	4,131	&	4,289	&	-2,633	&	-2,732	&	3,808	&	3,916	&	4,899	&	5,085	\\
\cline{1-9}																	
0,2	&	3,744	&	4,068	&	-2,406	&	-2,602	&	3,528	&	3,756	&	4,450	&	4,829	\\
\cline{1-9}																	
0,3	&	3,414	&	3,879	&	-2,209	&	-2,490	&	3,264	&	3,605	&	4,066	&	4,609	\\
\cline{1-9}																	
0,4	&	3,149	&	3,728	&	-2,048	&	-2,398  &	3,037	&	3,475	&	3,756	&	4,432	\\
\cline{1-9}																	
0,5	&	2,939	&	3,608	&	-1,916	&	-2,323	&	2,847	&	3,367	&	3,509	&	4,291	\\
\cline{1-9}																	
0,6	&	2,770	&	3,511	&	-1,809	&	-2,261	&	2,690	&	3,277	&	3,308	&	4,176	\\
\cline{1-9}																	
0,7	&	2,632	&	3,432	&	-1,721	&	-2,211	&	2,558	&	3,202	&	3,145	&	4,083	\\
\cline{1-9}																	
0,8	&	2,518	&	3,367	&	-1,647	&	-2,169	&	2,447	&	3,138	&	3,008	&	4,005	\\
\cline{1-9}																	
0,9	&	2,422	&	3,312	&	-1,584	&	-2,133	&	2,352	&	3,084	&	2,894	&	3,939	\\
\end{tabular}
}
\caption{Cantilever beam: numerical outcomes.}
\label{tab1}
\end{center}
\end{table*}


\begin{table*}[h!]
\begin{center}
\resizebox{1\textwidth}{!}{
\begin{tabular}{c|cc|cc|cc|cc}  
\multicolumn{1}{c}{$\lambda$}&\multicolumn{2}{c}{$u_{\bot}(0)\: [10^{-2}nm] $}&\multicolumn{2}{c}{$u_{\bot}(L)\:[10^{-2}nm]$}&\multicolumn{2}{c}{$\varphi(L)\: [10^{-3}]$}&\multicolumn{2}{c}{$u^{tot}_{max}(0) \:[10^{-2}nm] $} \\
$ $ & $ m = 0.3 $   &  $ m = 0.6  $ &   $ m = 0.3 $   &  $ m = 0.6  $  & $ m = 0.3 $   &  $ m = 0.6  $  & $ m = 0.3 $   &  $ m = 0.6  $  \\
\hline
\cline{1-9}
0,1	&	3,665	&	3,809	&	-3,527	&	-3,652&	-3,945&	-4,073		&	3,665	&	3,809\\
\cline{1-9}													
0,2	&	3,321	&	3,613	&	-3,196	&	-3,463&	-3,652	&	-3,905&	3,321	&	3,613	\\
\cline{1-9}													
0,3	&	3,030	&	3,447	&	-2,911	&	-3,300&	-3,383	&	-3,751	&	3,030	&	3,447\\
\cline{1-9}													
0,4	&	2,798	&	3,314	&	-2,683	&	-3,170&	-3,152	&	-3,619&	2,798	&	3,314	\\
\cline{1-9}													
0,5	&	2,613	&	3,208	&	-2,502	&	-3,067	&	-2,959	&	-3,509&	2,613	&	3,208	\\
\cline{1-9}													
0,6	&	2,464	&	3,123	&	-2,357	&	-2,984	&	-2,797	&	-3,417	&	2,464	&	3,123\\
\cline{1-9}													
0,7	&	2,342	&	3,054	&	-2,239	&	-2,916	&	-2,662	&	-3,339&	2,342	&	3,054	\\
\cline{1-9}													
0,8	&	2,241	&	2,996	&	-2,141	&	-2,860	&	-2,547	&	-3,274&	2,241	&	2,996	\\
\cline{1-9}													
0,9	&	2,156	&	2,947	&	-2,058	&	-2,813	&	-2,450	&	-3,218&	2,156	&	2,947	\\
\end{tabular}
}
\caption{Slider and roller supported beam: numerical outcomes.}
\label{tab2}
\end{center}
\end{table*}

\begin{table*}[h!]
\begin{center}
\resizebox{1\textwidth}{!}{
\begin{tabular}{c|cc|cc|c|c}  
\multicolumn{1}{c}{$\lambda$}&\multicolumn{2}{c}{$u_{\bot}(L)\: [10^{-2}nm] $}&\multicolumn{2}{c}{$\varphi(L)\: [10^{-3}]$}&\multicolumn{1}{c}{$u^{tot}_{max}(s^*) \:[10^{-2}nm] $}&\multicolumn{1}{c}{$u^{tot}_{max}(s^*) \:[10^{-2}nm] $} \\
$ $ & $ m = 0.3 $   &  $ m = 0.6  $ &   $ m = 0.3 $   &  $ m = 0.6  $  &   $ m = 0.3 $   &  $ m = 0.6  $  \\
\hline
\cline{1-7}
0,1	&	-1,036	&	-1,135	&	-1,658	&	-1,751	&	1,136	\:\:	$s^*$ = 5,892	&	1,241	\:\:	$s^*$ = 5,738	\\
\cline{1-7}																	
0,2	&	-0,815	&	-1,010	&	-1,474	&	-1,648	&	0,921	\:\:	$s^*$ = 6,041	&	1,117	\:\:	$s^*$ = 5,788	\\
\cline{1-7}																	
0,3	&	-0,667	&	-0,926	&	-1,330	&	-1,568	&	0,787	\:\:	$s^*$ = 6,063	&	1,041	\:\:	$s^*$ = 5,774	\\
\cline{1-7}																	
0,4	&	-0,574	&	-0,874	&	-1,221	&	-1,506	&	0,703	\:\:	$s^*$ = 6,042	&	0,993	\:\:	$s^*$ = 5,750	\\
\cline{1-7}																	
0,5	&	-0,515	&	-0,840	&	-1,137	&	-1,458	&	0,648	\:\:	$s^*$ = 6,009	&	0,962	\:\:	$s^*$ = 5,727	\\
\cline{1-7}																	
0,6	&	-0,477	&	-0,818	&	-1,072	&	-1,421	&	0,609	\:\:	$s^*$ = 5,974	&	0,940	\:\:	$s^*$ = 5,707	\\
\cline{1-7}																	
0,7	&	-0,450	&	-0,802	&	-1,020	&	-1,391	&	0,580	\:\:	$s^*$ = 5,940	&	0,923	\:\:	$s^*$ = 5,691	\\
\cline{1-7}																	
0,8	&	-0,431	&	-0,791	&	-0,977	&	-1,366	&	0,559	\:\:	$s^*$ = 5,910	&	0,911	\:\:	$s^*$ = 5,678	\\
\cline{1-7}																	
0,9	&	-0,418	&	-0,784	&	-0,942	&	-1,346	&	0,542	\:\:	$s^*$ = 5,884	&	0,902	\:\:	$s^*$ = 5,667	\\
\end{tabular}
}
\caption{Clamped and roller supported beam: numerical outcomes.}
\label{tab3}
\end{center}
\end{table*}


\section{Closing remarks}
\label{Concl}

The stress-driven mixture model of elasticity developed by \citet{BarrettaPhysE2018} for straight structures has been generalized in the present paper to model and assess size effects in small-scale curved stubby beams.
The relevant elastostatic problem has been preliminarily formulated by making recourse to Timoshenko kinematic theory, shown to be mathematically well-posed and analytically addressed by a simple and effective coordinate-free solution procedure. 

Selected case-studies of current interest in nano-mechanics have been studied and corresponding closed-form solutions have been detected by exploiting the aforementioned solution technique. 

Advantageously, the presented approach, driven by two parameters, has been proven to be able to simulate both softening and stiffening responses when compared with classical local structural behaviours.

Thus, the new methodology is technically appropriate for design and optimize the size-dependent nonlocal behaviour of a wide class of new-generation technological devices, such as Micro- and Nano-Electro-Mechanical Systems (M/NEMS) composed of small-scale curved beams.

\section*{Acknowledgments}
Financial supports from the MIUR in the framework of the Project PRIN 2017 (code 2017J4EAYB \emph{Multiscale Innovative Materials and Structures (MIMS)}; University of Naples Federico II Research Unit) is gratefully acknowledged.



\end{document}